\documentclass[a4paper,twocolumn,aps,superscriptaddress]{revtex4}

\usepackage{graphicx}
\usepackage{amsmath,amssymb}
\usepackage{bm}
\usepackage[usenames]{color}
\usepackage{subfigure}
\usepackage{epstopdf}

\definecolor{Nathanblue}{rgb}{0.96,0.24,0.00}

\usepackage[breaklinks=true,colorlinks=true,linkcolor=blue,urlcolor=blue,citecolor=blue]{hyperref}
\usepackage{graphicx,color}

\newcommand{\ahd}{\hat a^\dagger}
\newcommand{\ah}{\hat a}
\newcommand{\bhd}{\hat b^\dagger}
\newcommand{\bh}{\hat b}

\newcommand{\eq}[1]{(\ref{eq:#1})}
\newcommand{\eqname}[1]{\label{eq:#1}}
\newcommand{\rr}{\mathbf{r}}

\begin{document}

\title{Synthetic dimensions in integrated photonics: \\ From optical isolation to four-dimensional quantum Hall physics}

\author{Tomoki Ozawa}
\affiliation{INO-CNR BEC Center and Dipartimento di Fisica, Universit\`{a} di Trento, I-38123 Povo, Italy}
\author{Hannah M. Price}
\affiliation{INO-CNR BEC Center and Dipartimento di Fisica, Universit\`{a} di Trento, I-38123 Povo, Italy}
\author{Nathan Goldman}
\affiliation{CENOLI, Facult{\'e} des Sciences, Universit{\'e} Libre de Bruxelles (U.L.B.), B-1050 Brussels, Belgium}
\author{Oded Zilberberg}
\affiliation{Institute for Theoretical Physics, ETH Zurich, 8093 Z{\"u}rich, Switzerland}
\author{Iacopo Carusotto}
\affiliation{INO-CNR BEC Center and Dipartimento di Fisica, Universit\`{a} di Trento, I-38123 Povo, Italy}

\begin{abstract}

Recent technological advances in integrated photonics have spurred on the study of topological phenomena in engineered bosonic systems. Indeed, the controllability of silicon ring-resonator arrays has opened up new perspectives for building lattices for photons with topologically nontrivial bands and integrating them into photonic devices for practical applications. Here, we push these developments even further by exploiting the different modes of a silicon ring resonator as an extra dimension for photons. Tunneling along this synthetic dimension is implemented via an external time-dependent modulation that allows for the generation of engineered gauge fields. We show how this approach can be used to generate a variety of exciting topological phenomena in integrated photonics, ranging from a topologically-robust optical isolator in a spatially one-dimensional (1D) ring-resonator chain to a driven-dissipative analog of the 4D quantum Hall effect in a spatially 3D resonator lattice. Our proposal paves the way towards the use of topological effects in the design of novel photonic lattices supporting many frequency channels and displaying higher connectivities.

\end{abstract}

\maketitle

\section{Introduction}
\label{sec:introduction}

The study of topological systems originated in the setting of electronic condensed matter~\cite{TKNN, Hasan:2010, Qi:2011}. In these systems, the energy bands can be characterized by nontrivial topological invariants, which are robust under small perturbations. Importantly, these invariants can have many physical consequences such as quantized bulk responses and topologically protected edge physics, as has been experimentally investigated in a wide range of solid-state materials~\cite{Hasan:2010, Qi:2011}.

More recently, the introduction of topological concepts into photonics~\cite{HaldaneRaghu} has opened up many exciting avenues of research. Much of this activity has been focused on the experimental observation of topologically protected edge states in systems ranging from gyromagnetic photonic crystals in the microwave domain~\cite{Soljacic}, to Floquet topological insulators based on arrays of coupled waveguides~\cite{Rechtsman, Kraus}, to arrays of ring resonators in integrated silicon photonics~\cite{Hafezi}, and to engineered bianisotropic metamaterials~\cite{Slobozhanyuk}. In all these works, spatially periodic structures act as lattices for light which, in combination with an engineered synthetic gauge field, lead to two-dimensional (2D) photonic energy bands that can be characterized by a single topological invariant - the first Chern number. Going beyond two dimensions, the first experimental works on three-dimensional (3D) lattices have recently unveiled peculiar and intriguing topological features in their photonic band dispersion, such as Weyl points~\cite{Lu:2013,Lu:2015}. Furthermore, much interest has also been devoted to new protocols to measure the topological invariants in the bulk of the system as well as at its edges~\cite{Ozawa:2014,Bardyn:2014,Poshakinskiy:2015,Hu:2015}.

At the same time as this progress in photonics, the community of ultracold atomic gases has also been very active in the study of topological physics. To this end, artificial gauge fields were generated by dressing atomic lattices with suitable combinations of magnetic and optical fields~\cite{Dalibard2011,Goldman:2014bv}, such that neutral atoms behave like electrons under strong magnetic fields with measurable bulk (topological) responses~\cite{aidelsburger2013, miyake2013, jotzu2014,Aidelsburger:2015}. Very recently, exciting new perspectives have arisen from the combination of such engineered gauge fields with a synthetic dimension, where internal atomic degrees of freedom are exploited as an additional lattice dimension, as has been theoretically proposed~\cite{Boada2012, Celi:2014} and quickly experimentally realized in a 1D spatial geometry that ends up behaving as a 2D topological system~\cite{Mancini:2015,Stuhl:2015}. These advances have inspired us in a previous work to propose a method for the realization of 4D topological systems using atoms~\cite{Price:2015}. Indeed, such a conceptual shift towards synthetic dimensions holds great promise also in optics, as suggested by other recent proposals~\cite{Schmidt:2015, Luo:2015,Schleier, Fan}.

In this paper, we describe how the idea of a synthetic dimension can be exploited in an integrated photonics architecture, leading to engineered gauge fields and nontrivial band topologies in systems with one to three spatial dimensions. Our proposal relies on a ring-resonator array infrastructure where the different modes of each resonator act as the synthetic dimension. Coupling between different resonator modes is produced through an external time-dependent modulation of the dielectric properties of each cavity at a frequency equal to (or a multiple of) the free-spectral range. Additionally, this modulation can be tailored spatially such that each resonator experiences different (complex) coupling matrix elements; this can be used for engineering gauge fields and nontrivial band topologies. Importantly, our proposal is realizable with state-of-the-art silicon photonics technology, while the required modulation scheme can be implemented in an all-optical way. This allows one to control the synthetic gauge field with great flexibility and in a quickly reconfigurable way, just by changing the geometry of the strong modulating field.

Our scheme has important implications: (i) 1D chains of resonators coupled in this way behave as 2D topological systems with propagating edge states that may be exploited to provide on-chip optical isolation; and (ii) photonic topological lattice models can be constructed in dimensions higher than three dimensions. This may have dramatic applications in integrated photonic circuits with higher connectivity, where the mode index can be naturally associated with the different frequency channels of an optical signal.

\subsection*{Outline}

In Sec.~\ref{sec:synthetic}, we introduce the general strategy for generating a synthetic dimension in a ring resonator. We then illustrate our proposal for two example systems of experimental interest; first, in Sec.~\ref{sec:2D}, we show how to build a topologically protected optical isolator in a 1D chain of multimode ring-resonators. Second, in Sec.~\ref{sec:4D}, we show how a 3D resonator lattice could be used to probe a driven-dissipative analog of the 4D quantum Hall effect. In Sec.~\ref{sec:discussion}, we discuss our proposal in the context of earlier works in photonics, and finally, in Sec.~\ref{sec:conclu}, we discuss future perspectives offered by the introduction of synthetic dimensions into integrated photonics.

\section{Synthetic dimension}
\label{sec:synthetic}

In this section, we detail how the different modes of a silicon ring resonator can be treated as a discrete set of lattice sites along an additional synthetic dimension. After introducing this idea for a single resonator, we discuss how this approach can be naturally extended to an array of resonators to build effective higher-dimensional photonic lattices.

\subsection{The Single Resonator}
\label{sec:single}

We consider a single ring-resonator with radius $R$ and effective mode index $n_{\rm eff}$. Each optical mode is labeled by an integer $w$ corresponding to the angular momentum of the mode around the ring. The corresponding mode frequencies $\Omega_w$ are (almost) equispaced with a free spectral range $\Delta\Omega =2\pi c / (n_{\rm eff} R)$,
\begin{equation}
 \Omega_w=\Omega_{w_0}+\Delta \Omega\,(|w|-w_0) + \frac{D}{2} (|w|-w_0)^2 + \ldots,
\end{equation}
where the contribution of the frequency dispersion $D$, arising both from the refractive index dispersion of the cavity material and from the confinement geometry, is typically a small perturbation for modes sufficiently close to the reference mode $w_0 > 0$. 

\begin{figure}[tbp]
\begin{center}
\includegraphics[width=8.5cm]{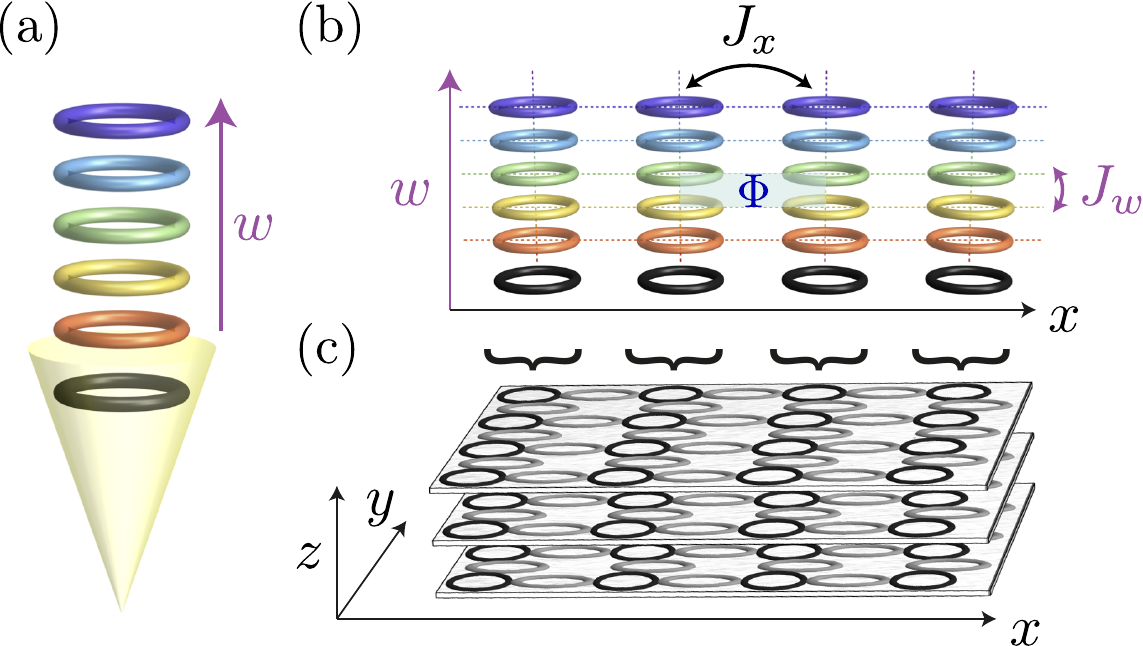}
\caption{(a) An illustration of the synthetic dimension scheme. The dielectric properties of a ring-resonator (black) are modulated by external light fields, leading to a linear coupling between the angular momentum eigenmodes of the resonator (colored rings represent different eigenfrequencies).
(b) An effective lattice made of a 1D array of resonators with coupling in a synthetic dimension. Here, $J_{x}$ denotes the tunneling amplitude along the real dimension by evanescent proximity coupling, and $J_w$ denotes the coupling in the synthetic dimension. Spatial variation of the dielectric modulation of each resonator makes the complex hopping amplitude $J_w$ spatially dependent. We can design this modulation so that the phase, $\Phi$, obtained by a photon hopping around a plaquette in the synthetic 2D lattice (shaded in blue) is uniform across the lattice. The resulting Hamiltonian is analogous to that of the Harper-Hofstadter model~\cite{Hofstadter:1976}, of a charged particle hopping on a 2D lattice in the presence of a uniform magnetic flux $\Phi$.
(c) A sketch of an effective 4D lattice composed of a 3D resonator lattice and one synthetic dimension. In the $xy$ plane the resonators are coupled via link resonators with directional couplers (gray rings). Making the position of the link resonators along the $y$ direction depend on $z$ engineers a gauge field and generates 2D quantum Hall physics in the $yz$ planes~\cite{Hafezi}. Similar to (b) the coupling of different eigenmodes along the synthetic dimension $w$ can be made to depend on $x$, leading to 2D quantum Hall systems in the $xw$ planes. The resulting 4D physics is that of the 4D integer quantum Hall effect on a hypercube~\cite{Price:2015,Kraus:2013}.}
\label{fig:overview}
\end{center}
\end{figure}

The key idea of our proposal is to view the different resonator modes as a discrete set of lattice sites along a synthetic dimension, $w$, as shown in Fig.~\ref{fig:overview}(a). Hopping processes along the synthetic dimension are obtained by linearly coupling modes with different values of $w$ through a suitable time-dependent modulation of the dielectric properties of the resonator. 
We propose to implement the time-dependent modulation using the material nonlinearity through the application of external light fields; a detailed discussion of this scheme is given in Appendix~\ref{sec:implementation}. The time-dependent modulation leads to an effective Hamiltonian of the form (taking $\hbar = 1$)
\begin{equation}
 H_{\rm mod }=-\sum_{w} \mathcal{J}_{\rr}(t)\,\ahd_{\rr,w+\eta} \ah_{\rr,w} + \textrm{H.c.},
 \eqname{Hmod}
\end{equation}
where the operator $\ahd_{\rr,w}$ creates a photon in mode $w$ of the resonator and the index $\rr$ refers to the spatial position of the resonator. We also introduce the hopping unit $\eta$, which is set by the details of the scheme used to realize the modulation of the dielectric properties of the resonator, as will be discussed in Appendix~\ref{sec:implementation}. For example, in Fig.~\ref{fig:overview}(a), the inter-mode coupling is depicted between modes that are adjacent in frequency, corresponding to $\eta=1$. The underlying physical process can be understood as a resonator photon in the $w$ mode being converted to the $w+\eta$ mode (and viceversa) by mixing with one or more quanta of the modulation field.

In order to obtain a static inter-mode coupling Hamiltonian, we assume that the modulation of the dielectric properties $\mathcal{J}_{\rr}(t)$ is monochromatic at a frequency close to resonance with the energy spacing between neighboring modes, $\Omega_{\rm mod}\approx \eta \Delta\Omega$ with a (complex) amplitude $\mathcal{J}^o_\rr$, i.e., the modulation takes the form $\mathcal{J}_{\rr}(t) = \mathcal{J}^o_\rr e^{-i\Omega_\mathrm{mod} t}$.
Moving to a rotating frame by
\begin{equation}
 \bh_{\rr, w}(t) \equiv \ah_{\rr, w}(t) e^{i[\Omega_{w_0}+ (w-w_0) \Omega_{\rm mod}/\eta]t },
 \label{eq:b}
\end{equation}
and focusing around $w\!\approx\!w_0$ we arrive at the time-independent Hamiltonian
\begin{align}
 &H_{\Omega} =\sum_{w} \left[ -\mathcal{J}^o_{\rr} \bhd_{\rr, w+\eta} \bh_{\rr, w} + \textrm{H.c.} \right. \label{H1site} \\
&+   \left. \left((\Delta \Omega -\Omega_{\rm mod}/\eta)\,(w-w_0)+\frac{D}{2}(w-w_0)^2 \right) \bhd_{\rr, w} \bh_{\rr, w}  \right]. \notag
\end{align}
The first line in $ H_{\Omega}$ is analogous to that of a particle hopping on a tight-binding lattice along the $w$ direction, while the second line contains additional terms acting as an on-site potential for each frequency eigenmode. Importantly, the tunneling matrix elements describing motion along the synthetic dimension, $J_w \!\equiv\! \mathcal{J}^o_{\rr}\!=\!\vert \mathcal{J}^o_{\rr}\vert \exp [i \theta (\rr)]$, can be spatially dependent and complex valued as we discuss further in the next sections. In the following, unless otherwise stated, we restrict ourselves to the simplest case where we can neglect the uniform force along $w$ due to the detuning $(\Delta \Omega -\Omega_{\rm mod}/\eta)$. We also assume that the harmonic potential, proportional to $D$, is negligible. We postpone the further investigation of these additional terms to future work. 

The Hamiltonian (\ref{H1site}) is one of the central results of our paper. This shows how a time-dependent modulation can effect a static inter-mode coupling in a silicon ring resonator in the rotating frame. Notably, from this result, it is justified to interpret the mode index $w$ as indexing an additional synthetic dimension. In the following, we show how this Hamiltonian can be combined with other building blocks to create topological lattices in integrated photonics. 

\subsection{Lattice of Resonators}
\label{sec:lattice}

The advantage of harnessing the internal-mode structure of a ring resonator becomes apparent when coupling several resonators spatially. Consider extending this approach to a $d_\mathrm{real}$-dimensional cubic lattice of multimode ring resonators. Due to the additional motion along the synthetic dimension, light passing through the lattice experiences a $d = d_\mathrm{real}+1$-dimensional lattice in which the lattice-site coordinates are given by the composite vector $(\rr,w)$. For closely spaced resonators, the overlap of optical modes of neighboring resonators provides a spatial tunneling process describable within the standard tight-binding Hamiltonian,
\begin{equation}
 H_J=-\sum_{\rr,j,w} J_j \,\bhd_{\rr+{\mathbf a}_j,w} \bh_{\rr,w} + \textrm{H.c.} ,\label{eq:hj}
 \end{equation}
where the sum over ${\rr}$ runs over all resonator positions, ${\mathbf a}_j$ is the lattice unit vector in the real-space $j$ direction (e.g., $j \in \{ x, y, z\}$ for a $d_\mathrm{real}=3$ lattice), and $J_{j}$ is the corresponding hopping amplitude that quickly decreases with the inter-resonator distance. Notably, energy conservation guarantees that the tunnel coupling is effective only between modes with the same $w$. As introduced in Sec.~\ref{sec:single}, inducing a time-dependent modulation of the dielectric properties generates a coupling along the $w$ direction within each resonator. Crucially, the complex phase factor associated with this coupling can be suitably chosen so as to engineer gauge fields, leading to topological effects in photonics as will be discussed in the following sections.

To model a realistic experiment, we must include the inherent driving and dissipation present in a lattice of ring resonators. In such systems, light continuously leaks out from the resonators; we model these losses with a position- and mode-dependent loss rate $\gamma_{\rr,w}$. To balance the losses, light is continuously injected into the system through a monochromatic driving field. It can be shown that the cavity field expectation value $\beta_{\rr, w}(t) \equiv \langle \hat{b}_{\rr,w}(t)\rangle$ then evolves under the effects of the driving, dissipation and combined Hamiltonian $H=H_J + \sum_\mathbf{r} H_\Omega$ [c.f. Eqs.~(\ref{H1site}-\ref{eq:hj})] according to the equation of motion~\cite{Ozawa:2014}:
\begin{align}
	i\frac{\partial}{\partial t} \beta_{\rr,w} &= -\sum_j J_j (\beta_{\rr-{\mathbf a}_j,w}+\beta_{\rr+{\mathbf a}_j,w}) - \mathcal{J}^o_{\rr} \beta_{\rr,w-\eta}  \nonumber \\
	&  -[\mathcal{J}^o_{\rr}]^* \beta_{\rr,w+\eta} - i\gamma_{\rr,w} {\beta}_{\rr,w} + f_{\rr,w}(t),
 \label{driveneq}
\end{align}
where $f_{\mathbf{r},w}(t)$ is the driving term as seen in the rotating frame defined by (\ref{eq:b}).
In the following we shall focus on the case where $f_{\mathbf{r},w}(t)\propto e^{-i\Omega_{\rm drive} t}$ is monochromatic at a driving frequency $\Omega_{\rm drive}$. In the non-rotating frame, this corresponds to the driving frequency of $\Omega_w + \Omega_{\rm drive}$. Under the assumption that $\Omega_{\rm drive}\ll \Delta \Omega$, the driving is effectively restricted to one mode, $w$, only. All the simulations in this paper are performed with the driving field strength equal to the hopping: $|f_{\mathbf{r},w}(t)| = J$ for the driven site. Experiments on a sufficiently long time-scale probe the steady-state solution of the field, where the excitation has spread over many cavities and many modes. Each cavity mode $w^\prime$ oscillates, in the non-rotating frame, at a frequency $\Omega_{w^\prime} + \Omega_{\rm drive}$ in the steady state. The different supermodes of the many-resonator system can be selectively addressed by varying the position and frequency of the driving field~\cite{carusotto_review}. In particular, as we now discuss in turn in the context of topological band structures, it is possible to excite either mainly the bulk bands themselves~\cite{Ozawa:2014} or the topological gapless edge modes located in the bulk band gaps.

Before entering into the discussion of some remarkable effects that can be anticipated for the proposed set-up, it is important to conclude this general presentation by briefly commenting on the strength of the hopping in the synthetic dimension that is needed to probe the topological physics. In order to have sizable interband gaps that are able to protect the topological features, we need hopping in the real and synthetic dimensions to be of the same order and, most importantly, significantly larger than the cavity losses $\gamma$. Along the real dimension the tunnel amplitude is controlled by the spatial separation of neighboring cavities and there is no difficulty in satisfying this condition, as is done in recent experiments~\cite{Hafezi}. Along the synthetic dimension, this condition reads $|\mathcal{J}_\mathbf{r}^0| > \gamma$, and can be reformulated in terms of the cavity $Q$ factor $Q=\Omega_{w_0}/\gamma $ as $|\mathcal{J}_\mathbf{r}^0| / \Omega_{w_0} > 1/Q$. As we discuss in detail in Appendix~\ref{sec:implementation}, the quantity $|\mathcal{J}_\mathbf{r}^0| / \Omega_{w_0}$ is essentially the dielectric modulation one is able to generate in the material, and the condition $|\mathcal{J}_\mathbf{r}^0| / \Omega_{w_0} > 1/Q$ is practically achievable in high-quality resonators for which $Q$ factors as high as $10^5$ can be achieved~\cite{Hafezi}. For simplicity, in the following we will assume equal hopping amplitude for all the directions: $J \equiv J_j = |\mathcal{J}_\mathbf{r}^0|$.

\section{Topological Effects in a 1D Chain of Ring Resonators}
\label{sec:2D}

As introduced in the previous section, we can exploit the different modes of ring resonators to realize a higher-dimensional lattice in a resonator array. We now demonstrate the power of this approach for a chain of multimode resonators in which the light experiences an effective $(1+1)$-dimensional lattice with one synthetic dimension and one real spatial dimension, chosen here along $x$ as shown in Fig.~\ref{fig:overview}(b). 

Importantly, we show that this effective lattice can be used to simulate the 2D Harper-Hofstadter model, a seminal lattice model for the realization of the 2D quantum Hall effect. We then briefly review the basic properties of this model, emphasizing the interesting topological characteristics of its bulk energy bands and edge states. Then we discuss how the topology of both the bulk bands and the edge states may be probed in a driven-dissipative photonics experiment. While a measurement of the bulk topological invariant would be a remarkable realization of an analog optical quantum Hall effect, the topological edge states of this system may prove to be even more interesting as they may have important applications for optical isolation in integrated photonics.

\subsection{Simulating the Harper-Hofstader model by engineering a gauge field in the synthetic dimension}
\label{sec:2DA}

As introduced by Eq.~(\ref{H1site}), different cavity modes can be coupled in a ring resonator by temporally modulating the dielectric properties of the cavity. As this equation implies and as discussed further in Appendix~\ref{sec:implementation}, this modulation can be designed to depend on the resonator position. If we choose the modulation to have the explicit spatial dependence $\mathcal{J}_x^o = |\mathcal{J}_x^o| e^{ik_x x}$, the full Hamiltonian of the resonator lattice $H=  H_J + \sum_\mathbf{r} H_{\Omega}$ becomes
\begin{align}
	H =
	-\sum_{x,w} \left( J_x \hat{b}_{x+a,w}^\dagger \hat{b}_{x,w} + |\mathcal{J}_{x}^0| e^{ik_x x}\hat{b}_{x,w+\eta}^\dagger \hat{b}_{x,w} + \text{H.c.}\right) . \label{eq:ham2d}
\end{align}
This is the analog of the 2D (bosonic) Harper-Hofstadter model, describing a charged particle hopping on a 2D tight-binding square lattice in the presence of a perpendicular uniform magnetic field~\cite{Hofstadter:1976}. In this analogy, the effective $(1+1)$-dimensional lattice in the $xw$ plane in Eq.~(\ref{eq:ham2d}) can be recognized as the 2D square lattice of the Harper-Hofstadter model. Similarly, the complex hopping phase factor $e^{ik_x x}$ in Eq.~(\ref{eq:ham2d}) plays the role of the Peierls phase factor of a charged particle hopping in a magnetic field. 

When light hops around a single plaquette of our effective $(1+1)$-dimensional lattice, the light gains a total phase, which is equivalent to the magnetic flux piercing the plaquette. 
This is indicated in Fig.~\ref{fig:overview}(b), where a single plaquette of the lattice is shaded in blue, and the magnetic flux is denoted by $\Phi$. When the resonators are spaced by a distance $a$ along $x$, this enclosed magnetic flux is equal to $2\pi \Phi = \eta a eB = k_x a$ per plaquette, where $B$ is the artificial magnetic field, $e$ is the fictitious charge in the artificial magnetic field, and $\eta a$ is the area of the plaquette in the $xw$ plane. As we can see, the strength of this engineered magnetic field is  easily tuned through the wave vector $k_x$ of the dielectric modulation. This wave vector is in turn controlled externally via the optical fields that induce the dielectric modulation as will be discussed in Appendix~\ref{sec:implementation}.

\subsection{Introduction to the 2D Harper-Hofstadter model}

The Harper-Hofstadter model has important spectral and topological properties~\cite{Hofstadter:1976}, which we now briefly review. When the flux per plaquette $\Phi$ is rational and takes the form $\Phi = p/q$, where $p$ and $q$ are mutually prime integers, the energy spectrum of the Harper-Hofstadter model consists of $q$ bands. Each of the $q$ bands is topologically nontrivial, and characterized by a nonzero first Chern number. The non-zero first Chern number of bands implies that when the system is finite, namely, when the system has edges, topologically protected states localized at the edges exist at energies, which fall in the band gaps~\cite{Hasan:2010,Qi:2011}.

\begin{figure}[htbp]
\begin{center}
\includegraphics[width=8.5cm]{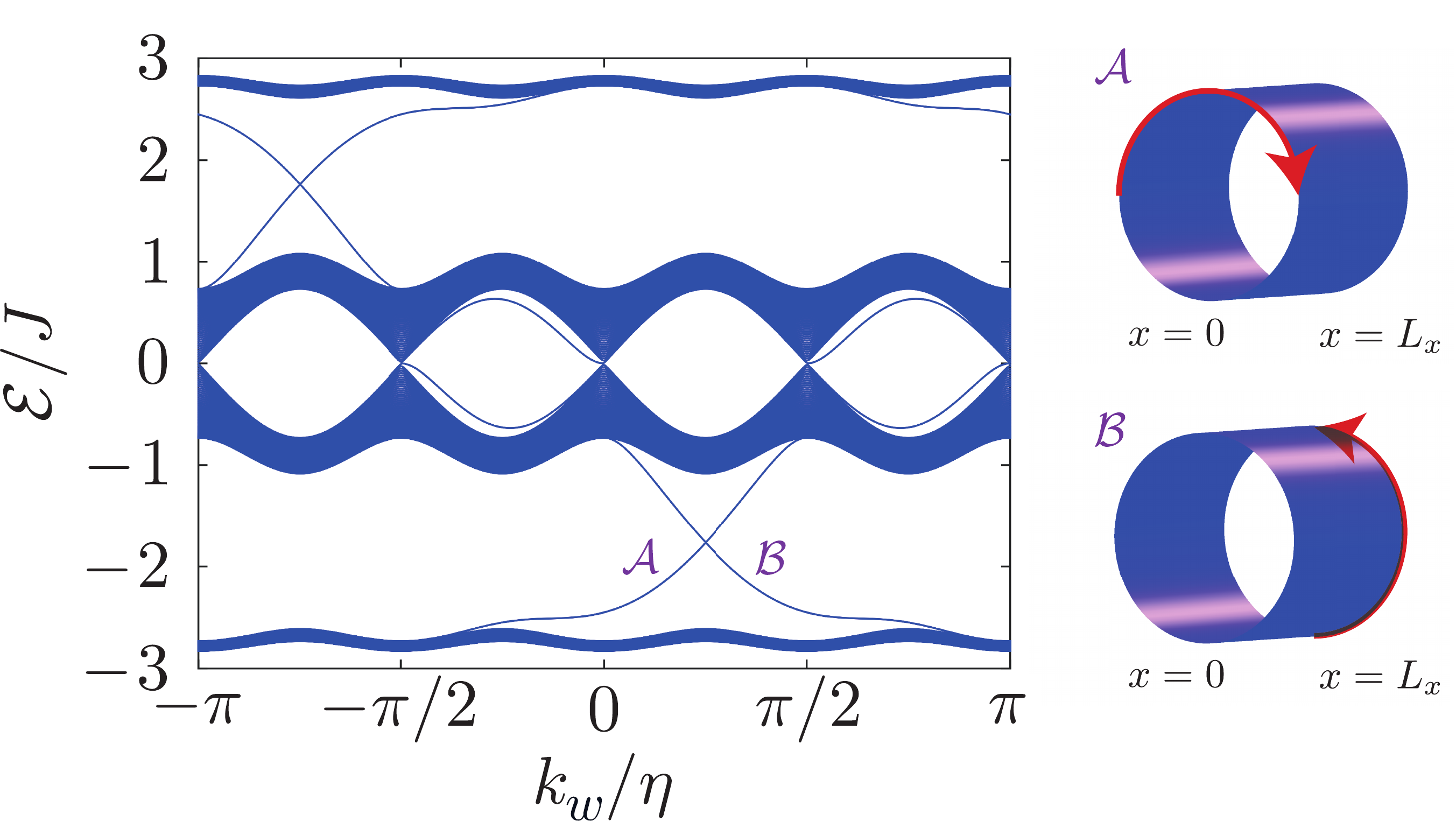}
\caption{(Left) The energy spectrum $\mathcal{E}(k_w)$ of the Harper-Hofstadter model with $\Phi = 1/4$. We take the $w$ direction to be periodic, and the $x$ direction to have a finite width $L_x$. There are four bands, with two bands touching at the zero energy. Two states inside each band gap are chiral edge states, one localized at one edge and the other localized at the other edge. (Right) Schematic representation of the edge states labeled $\mathcal{A}$ and $\mathcal{B}$ in the left figure. The edge state $\mathcal{A}$ is localized at the edge $x = 0$, and the other edge state $\mathcal{B}$ is localized at the edge $x = L_x$. These edge states in the lower gap would propagate in the counter-clockwise direction in a planar geometry.}
\label{fig:spectrum2dedge}
\end{center}
\end{figure}

In Fig.~\ref{fig:spectrum2dedge}, we plot the energy spectrum $\mathcal{E}(k_w)$ of the system as a function of the momentum in the $w$ direction for $\Phi = 1/4$.
Assuming that the synthetic dimension is large enough to neglect edge effects in that direction, we artificially close the system along $w$ by applying periodic boundary conditions. In the $x$ direction we assume that there is a finite number of sites with open boundary conditions at the edges; this generates an artificial cylinder geometry. One can observe four bulk bands, two of them touching at zero energy, as well as states traversing the gaps between the bulk bands. The two states inside each gap are localized at opposite edges of the cylinder. The edge modes located within each bulk gap are associated with a well-defined chirality (note that opposite edges of the cylinder have opposite orientations, so that chirality is indeed respected). The fact that each gap contains only states with a definite chirality shows that these edge states are topologically protected from back scattering when disorder is present at the edge. The cylinder analysis described here straightforwardly applies to realistic (planar) geometries.

\subsection{2D quantum Hall effect in a 1D resonator chain}
\label{sec:2Dbulk}

The first Chern number of a Harper-Hofstadter energy band can have direct consequences in the steady-state distribution of photons in the driven-dissipative system that we study here. This bulk topological invariant can be measured via an optical quantum Hall effect using the scheme proposed in Ref.~\cite{Ozawa:2014}. To observe this effect, we drive a single resonator in the middle of the chain at a frequency resonant with a bulk band of our 2D synthetic lattice. We then apply a uniform frequency gradient in the synthetic dimension, as described by the contribution from $\Delta \Omega - \Omega_\mathrm{mod}/\eta$ in Eq.~(\ref{H1site}). This term acts as a uniform synthetic electric field along the synthetic dimension. From our numerical simulations, we measure the center of mass of the steady-state photon distribution; the center of mass is displaced from the driven resonator due to the Lorentz force in the synthetic 2D plane. 

The center-of-mass displacement can be used as a measurement of the first Chern number of an isolated band, provided that the bandwidth $\Delta \mathcal{E}_{\rm band}$ is much smaller than the energy gap to adjacent bands $\Delta \mathcal{E}_{\rm gap}$. By setting the driving frequency on resonance with the band, it is possible to get the required (approximately) uniform population of a single band, provided that the loss rate satisfies $\Delta \mathcal{E}_{\rm band}<\gamma<\Delta \mathcal{E}_{\rm gap}$; then the corresponding displacement of the center of mass obeys~\cite{Ozawa:2014}
\begin{align}
	\langle x \rangle
	=
	-q\nu_1 (\eta a) \frac{eE_w}{2\pi \gamma} + \mathcal{O}(\gamma^0), \label{2dphotonformula}
\end{align}
where $E_w$ is the synthetic electric field applied in the synthetic dimension, $q$ is defined via $\Phi = p/q$ where $p$ and $q$ are coprime integers, and $\nu_1$ is the first Chern number of the occupied band. This is an analog of the integer quantum Hall effect in driven-dissipative systems. The second term, which is independent of $\gamma$, is a correction to the first term when $\gamma$ is small but larger than the bandwidth.
Hence, by measuring the displacement $\langle x \rangle$, one can estimate the first Chern number using the formula (\ref{2dphotonformula}).

\begin{figure}[htbp]
\begin{center}
\includegraphics[width=8.5cm]{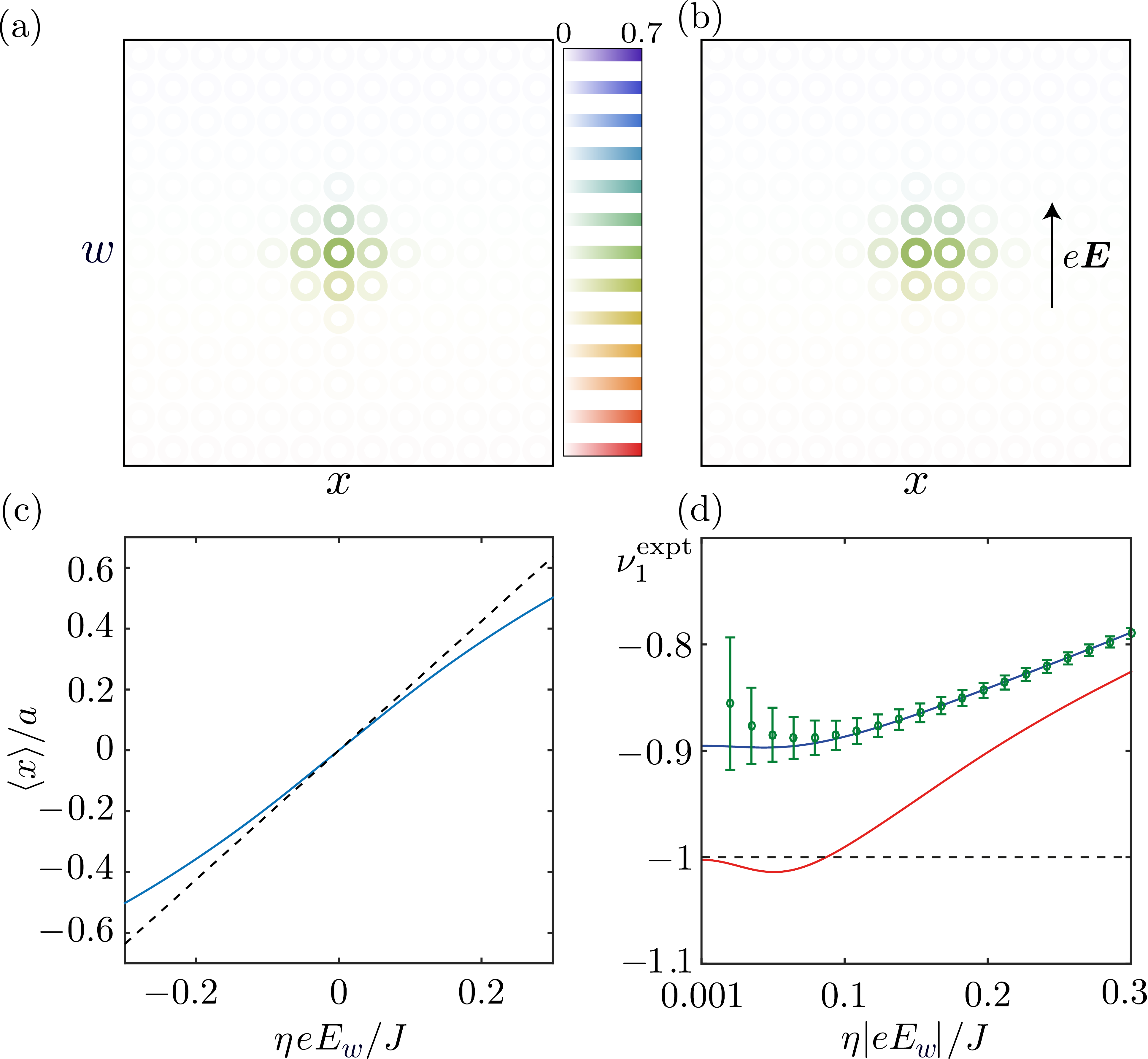}
\caption{(a), (b) The steady-state photon distribution $\beta_{x,w}$ in (\ref{driveneq}) for the Harper-Hofstadter model with $\Phi = 1/4$ when the central resonator is driven. The synthetic electric field $\eta eE_w$ is zero for (a), and $\eta eE_w = 0.3J$ for (b). The driving frequency is tuned to $-2.7 J$ within the lowest-energy band, and the uniform loss is $\gamma = 0.3J$, which is large enough to cover the lowest band. Different colors schematically represent that different sites in the synthetic dimension correspond to different frequencies. More opaque circles represent higher intensity; color bars shown in (a), also applicable to (b), show the value of the steady-state photon distribution calculated at each lattice site. (c) The numerically calculated displacement $\langle x\rangle$ (solid line) and the analytical prediction (dashed line) as a function of the electric field $E_w$. (d) The estimated Chern number, as a function of the nonzero force $|e E_w|$. The upper (blue) solid line is the estimate from Eq.~(\ref{2dphotonformula}) without the subleading term $\mathcal{O}(\gamma^0)$, and the points with error bars are estimated Chern numbers for systems with $\pm 10$\% random fluctuations on the loss rate $\gamma_{\mathbf{r},w}$, averaged over 20 different realizations. The lower (red) solid line is the estimate taking into account the term $\mathcal{O}(\gamma^0)$.  When the force $E_w$ is small, the estimated Chern number is close to the true value of $-1$, which is indicated by a dashed line.}
\label{fig:replot}
\end{center}
\end{figure}

In Figs.~\ref{fig:replot}(a) and~\ref{fig:replot}(b), we show the steady-state photon distribution without and with the uniform frequency gradient. We have solved Eq.~(\ref{driveneq}) for the Hamiltonian~(\ref{eq:ham2d}) with the driving field located at the center of the synthetic 2D system, and obtained the steady-state solution. We have chosen $\Phi = 1/4$ with $\Omega_\mathrm{drive} = -2.7J$ resonant with the lowest band. We have assumed a uniform loss rate of $\gamma = 0.3J$, which is large enough to cover the lowest band, whose bandwidth is $\sim 0.22J$. An approximately uniform loss rate in the synthetic dimension can be achieved by using the mode index $w_0 \gg 1$. This loss is also chosen to be smaller than the band gap to the upper band which is $\sim 1.53J$. As seen in Figs.~\ref{fig:replot}(a) and~\ref{fig:replot}(b), when there is no force in the synthetic dimension, the center of mass does not move from the driven site. On the other hand, in the presence of a force along the synthetic dimension, the center of mass is displaced in the real dimension. 

In Fig.~\ref{fig:replot}(c), we plot the displacement $\langle x\rangle$ as a function of the synthetic electric field $E_w$, as well as the prediction from the formula (\ref{2dphotonformula}) without the term $\mathcal{O}(\gamma^0)$. The estimated first Chern number from the formula (\ref{2dphotonformula}) as a function of $e E_w$ is plotted in Fig.~\ref{fig:replot}(d); the upper solid line corresponds to the estimate when we do not take into account the term $\mathcal{O}(\gamma^0)$, while the lower solid line is the estimate when we use two different values of the loss $\gamma = 0.3J$ and $0.31J$ to calculate $\langle x\rangle$ and subtract one from the other to get rid of the contribution from the term $\mathcal{O}(\gamma^0)$.
The estimated Chern number is close to the true value of $-1$ when $|e E_w|$ is small; this estimate is further improved when we eliminate the term $\mathcal{O}(\gamma^0)$.

In order to assess the robustness of our proposed measurement, we have repeated our simulations of the former scheme in the presence of random on-site disorder of the loss rate $\gamma_{{\mathbf r},w}$. As one can see in Fig.~\ref{fig:replot}(d), the statistical error bars due to averaging over disorder affect our proposal only moderately as long as the synthetic electric field is larger than the fluctuation in the loss rate.

As first numerically found but not specifically discussed in Ref.~\cite{Ozawa:2014}, it is interesting to note that a small displacement of the photon intensity distribution also occurs in the direction $w$ of the synthetic electric field. This displacement is not predicted by the ordinary quantum Hall theory in electronic systems, which predict instead a vanishing longitudinal electrical conductivity. Its discussion requires a more sophisticated analysis of the steady-state solution of the driven-dissipative wave equation in the presence of the synthetic electric field, which will be given in a future work.

In Sec.~\ref{sec:4D} we apply a similar idea to the 4D case, where this method can be used to estimate the second Chern number, a distinct topological invariant that characterizes the topology of systems in four dimensions.

\subsection{Topological edge states and optical isolation}

One of the most remarkable features of topological systems is the existence of robust edge physics. As shown in Fig.~\ref{fig:spectrum2dedge}, the Harper-Hofstadter model, in the absence of driving and dissipation, hosts one-way-propagating edge states in the energy gaps between the topological bulk bands. As these states are topologically protected, they cannot be destroyed by any perturbations, such as weak impurities or disorder, which do not close an energy band gap. We now investigate how the edge states of the effective 2D lattice can be studied and exploited in a 1D chain of temporally modulated multimode resonators.

Although the spatial edge along the $x$ direction is sharply defined by the ends of the resonator chain, the synthetic dimension can in principle consist of a semi-infinite number of modes, e.g. all modes with $w>0$. Using cavities with $w_0 \gg 1$, we typically work in a regime where this edge at $w = 0$ is not accessible. Without a hard boundary in the synthetic dimension, light in the topological edge state will propagate in the frequency direction but will not propagate significantly along the resonator chain. 

\begin{figure}[htbp]
\begin{center}
\includegraphics[width=8.5cm]{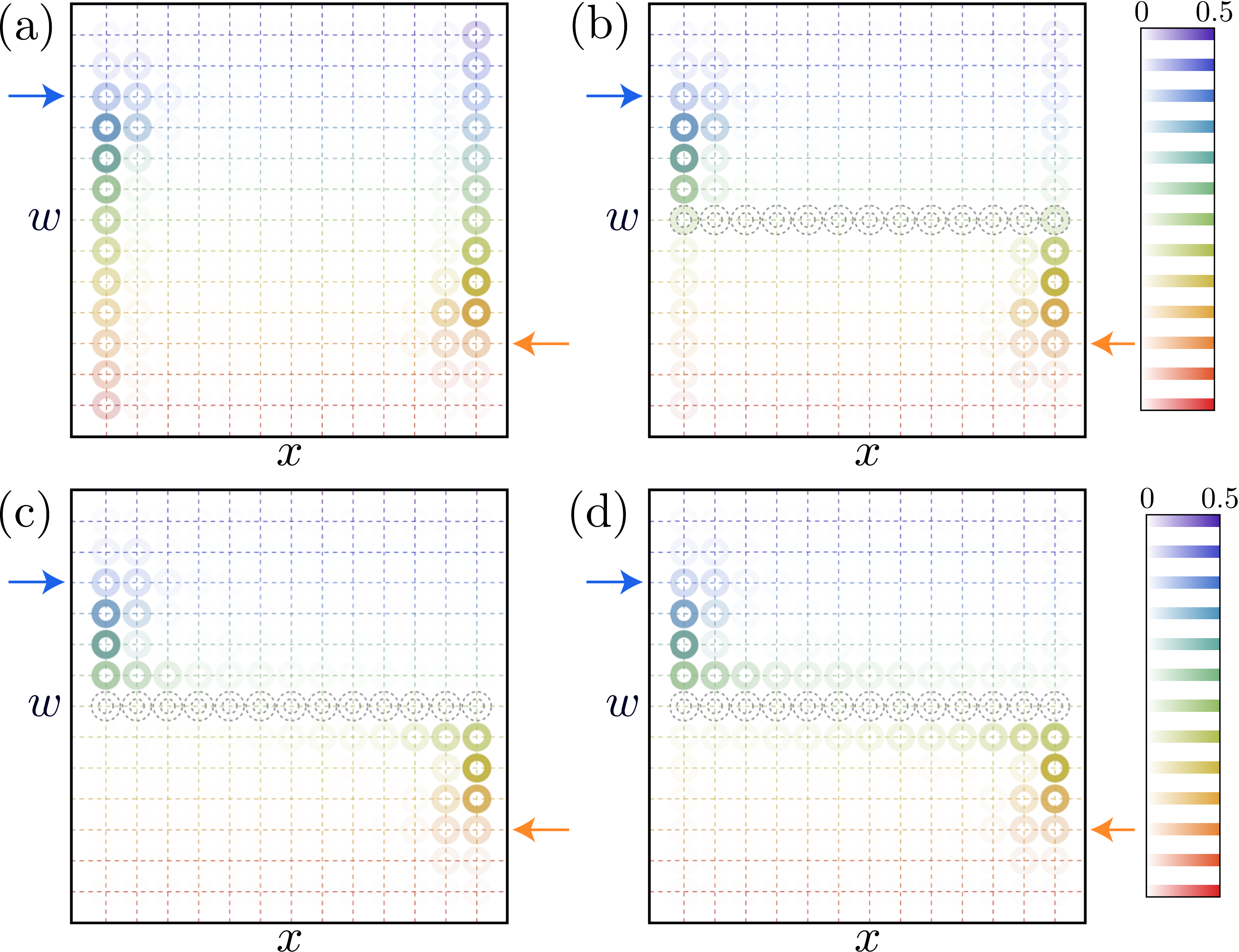}
\caption{Simulation of the steady-state photon-field distribution for $\Phi = 1/4$. Light is pumped into the system with a driving frequency $\Omega_{\mathrm{drive}} = -2J$,  chosen in the bulk band gaps of the 2D Harper-Hofstadter model (see Fig.~\ref{fig:spectrum2dedge}). The uniform loss rate is $\gamma = 0.1J$. The pumped modes and positions are indicated by arrows, and the central modes (indicated by black circles) have losses equal to (a) $\gamma$ (i.e., no extra losses), (b) $10\gamma$, (c) $100\gamma$, and (d) $1000\gamma$. The color bars show the value of the steady-state photon distribution calculated at each lattice site. In (a) where there are no extra losses in the central row, the light remains spatially localized at either end of the chain, while changing its frequency depending on the chirality of the edge state driven. When the loss of the central row becomes very large, as in (d), the lossy mode acts like an artificial edge in the synthetic dimension. Depending on the frequency of the pumped mode, light propagates either to the left or the right. The unidirectional propagation of these edge states can be exploited to provide topological optical isolation (see also Fig.~\ref{fig:transmission}).}
\label{fig:2dedge}
\end{center}
\end{figure}

The frequency conversion of light by the topological edge states is shown in Fig.~\ref{fig:2dedge}(a), where we present numerical simulations of the steady-state photon-field distribution for $\Phi = 1/4$, where the driving frequency $\Omega_{\mathrm{drive}} = -2J$ is chosen in the bulk band gaps of the 2D Harper-Hofstadter model (Fig.~\ref{fig:spectrum2dedge}), and the boundary resonators are pumped. The loss-rate is taken to be uniform and equal to $\gamma = 0.1J$. As can be seen, if the lattice is pumped at the left-most resonator, the light cascades down in frequency. While, if instead the lattice is pumped at the right-most resonator, the light increases in frequency due to the chirality of the edge state. This behavior can be reversed by changing the driving frequency to a different bulk bandgap where edge states have opposite chirality.

We note that there may also be a smooth harmonic confining (expelling) potential for $D>0$ ($D<0$) from the frequency dispersion of the cavity modes [Eq.~(\ref{H1site})]. This will not be discussed further here, except to point out that this situation is opposite from that in ultracold atomic gases, where the spatial dimension is usually harmonically trapped while the synthetic dimension has sharp edges~\cite{Celi:2014,Mancini:2015,Stuhl:2015}. 

To create a sharp boundary in the synthetic dimension, we exploit the intrinsic driven-dissipative nature of the photonic system. If impurities are inserted into the resonators, which have a narrow absorption line tuned close to resonance with a specific mode, this mode becomes much more lossy than the others. If these extra losses are strong enough, the Zeno effect~\cite{Zeno} sets in and makes the absorbing sites act as impenetrable and perfectly reflecting sites~\footnote{Although the Zeno effect is typically associated with quantum mechanics, it is a much more general feature of wave mechanics. A well-known manifestation of it in a classical electrodynamics framework is the almost perfectly lossless reflectivity at the surface between vacuum and a strongly absorbing medium, independently of the value of the real part of refractive index of the latter~\cite{Jackson}.}. This dramatically affects the photon steady state, as can be seen in Figs.~\ref{fig:2dedge}(b)-\ref{fig:2dedge}(d): light reaching this lossy mode from higher frequencies will avoid it by traveling to the right, while light reaching this mode from lower frequencies will travel to the left. The presence of strong loss therefore engineers an effectively lossless edge in the synthetic dimension. Note that this technique of creating edges in the synthetic dimension is a very flexible one, as, in principle, the edge can be defined along an arbitrary direction in the $xw$ plane, by making different modes lossy in different resonators.

\begin{figure}[htbp]
\begin{center}
\includegraphics[width=8.5cm]{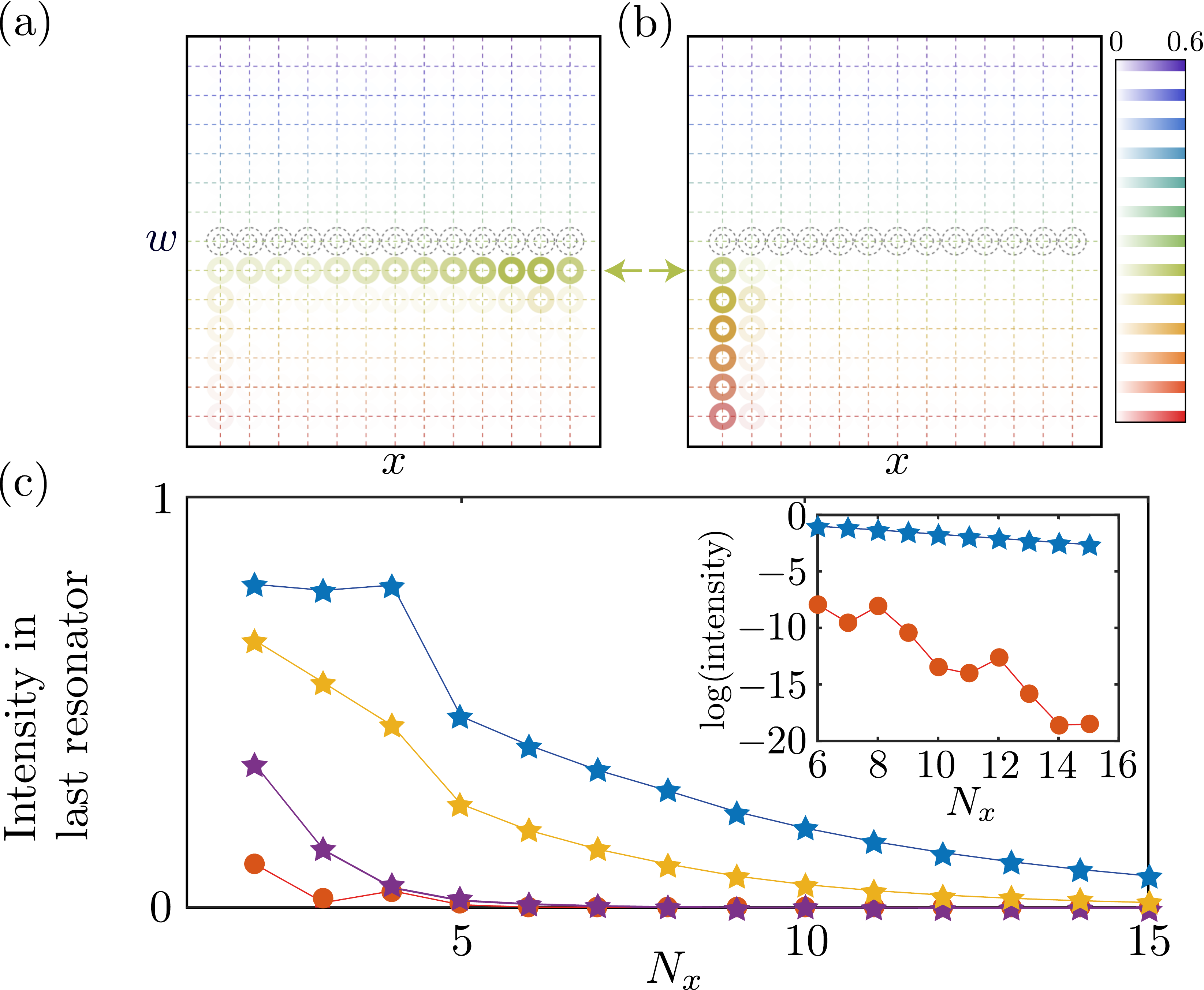}
\caption{Demonstration of how the synthetic two-dimensional lattice can be used as an optical isolator. (a), (b) Steady-state photon configurations when (a) the rightmost and (b) the leftmost sites are driven (as indicated by the arrows). The loss rate of all modes is taken to be uniform and equal to $\gamma = 0.1J$, except for the mode directly above the driven one [indicated as dashed circles in (a) and (b)] whose loss is set to be 1000 times larger. The color bars show the value of the steady-state photon distribution calculated at each lattice site. (c) Plot of the steady-state intensity at one end of the chain, when the other end is driven, as a function of the length of the chain $N_x$. The three topmost curves (blue, yellow, and purple stars) correspond to a drive located on the rightmost resonator. These three curves differ by the value of the extra losses of the mode defining the topological edge, which is equal, from top to bottom, to $1000\gamma$, $100\gamma$, and $10\gamma$. The bottom curve (red circles) corresponds to a drive located on the leftmost resonator with an extra loss rate of $1000\gamma$: the almost vanishing value of the intensity on the rightmost exit resonator shows that the system indeed acts as a good optical isolator. The ratio between the stars and circles quantitatively characterizes the efficiency of the optical isolation. The log-scale zoom of the curves for the extra loss rate of $1000\gamma$ in the region $N_x=6$ to $15$ shown in the inset further illustrates that there is always a large difference in transmission in opposite directions.}
\label{fig:transmission}
\end{center}
\end{figure}

The creation and control of topological edge states in a 1D chain of resonators can have direct applications as robust optical isolators for integrated photonics. In an optical isolator, light is transmitted in only one direction~\cite{Jalas:2013}; such devices are important in photonics circuits to protect and integrate different components on chip. As we see in Fig.~\ref{fig:2dedge}, depending on the frequency of the incident light compared to the lossy frequency, the light can only travel along the 1D resonator chain in one direction. For example, consider light injected at a frequency below the lossy mode, as described in Figs.~\ref{fig:transmission}(a) and~\ref{fig:transmission}(b). If this light is injected from the right side of the resonator chain, it propagates along the chain to the left as shown in Fig.~\ref{fig:transmission}(a). If instead, the same frequency light is injected from the left side of the resonator chain as shown in Fig.~\ref{fig:transmission}(b), it would cascade down in frequency, due to the chirality of the topological edge state. As there is typically no lower edge in the synthetic dimension, this light would not propagate along the chain. As the edge states are topologically robust, the operation of this system as an optical isolator should be robust and unaffected by weak backscattering from fabrication imperfections or surface roughness.

In order to quantitatively assess the efficiency of this system as an optical isolator, we inject light from one end of the resonator chain at the mode right below the lossy mode, and observe the transmitted intensity from the resonator at the other end. In our model, the emission from a site is simply proportional to the light intensity present on it. A plot of the transmitted intensity as a function of the length of the resonator chain $N_x$ is plotted in Fig.~\ref{fig:transmission}(c). As we can see, when we drive the leftmost resonator, there is almost no emission from the rightmost resonator, whereas when we drive the rightmost resonator, the leftmost resonator emits a significant amount of light.

As the extra loss rate of the lossy mode becomes larger, the transmission from the leftmost resonator becomes stronger thanks to the improved performance of the Zeno reflection. We have verified that, with the strongest extra loss rate shown in the figure, the lossy mode essentially acts as a sharp lossless edge. This statement is quantitatively confirmed by the fact that, for sufficiently long chains, the transmission decays as a function of chain
length according to an exponential law of characteristic absorption length $v_{\rm gr}/2\gamma$ proportional to the edge state group velocity $v_{\rm gr}$ and inversely proportional to the intrinsic loss rate $\gamma$, independently of the precise value of the (very strong) extra loss rate.

Deviations from the exponential law are instead clearly visible in shorter chains for the strongest values of the extra loss rate, see for instance the plateau-like feature followed by a marked kink at $N_x = 4$ on the top starred curve in Fig.~\ref{fig:transmission}(c). Several effects concur in introducing such short-distance corrections to the intensity profile. Non-resonant excitation of the bulk band states gives a spatially localized contribution to the intensity profile around the pumped site. In addition to this non-resonant excitation effect, a general feature of chiral modes resonantly excited by a spatially localized pump is that the spatial intensity profile is peaked on the sites just next to the driven site in the propagation direction rather than on the driven site itself.

This spatial displacement is clearly visible in the spatial intensity profiles shown in Fig.~\ref{fig:2dedge} as well as in Figs.~\ref{fig:transmission}(a),~\ref{fig:transmission}(b) and is qualitatively explained by the analytical model presented in Appendix~\ref{sec:analyticalmodel}. In Fig.~\ref{fig:transmission}(c) it contributes to the plateau-like feature on the top starred curve. For the two lower starred curves, the smaller extra losses at the lossy mode lead to a softer (not sharp) edge, whose effect can be modeled by a reduced effective absorption length. For such shorter absorption lengths, in agreement with the analytical model in Appendix~\ref{sec:analyticalmodel}, the amount of the spatial shift decreases and the intensity profile gets more and more peaked in the close vicinity of the driven site, eventually recovering the smooth and monotonous decay shown by the lower two starred curves of Fig.~\ref{fig:transmission}(c).

On the other hand, the transmission from the left to right shown by the circles in Fig.~\ref{fig:transmission}(c) is suppressed by several orders of magnitude, which confirms that, thanks to the existence of chiral topological edge states, our system indeed acts as an optical isolator. The completely different nature of the transmission in the two directions is apparent in the inset of Fig.~\ref{fig:transmission}(c), where the $N_x$ decay of the leftward (allowed) direction is much slower than the one in the (forbidden) rightward direction; in the former case the decay is in fact due to absorption, while in the latter case propagation occurs only on short distances via the evanescent tail of non-resonantly excited modes.

In a practical device, the system length must be tuned according to a trade-off between the optical isolation ratio and the actual transmittivity of the system. Most remarkably, the simulation in Fig.~\ref{fig:transmission} shows that a good isolation is already obtained with just two sites in the $x$ direction, a regime where the edge state has not yet fully developed~\cite{Hu:2015}. Of course, this result requires that there is no spurious light propagation across the system that does not go through the considered resonator modes. A careful design of the structure should, however, allow for the suppression of this spurious extrinsic propagation channel to arbitrarily low levels.

A possible limitation of our system as an optical isolator device is the bandwidth on which it is able to operate. As optical isolation is based on an edge state, its bandwidth is limited by the spectral width of the crossed energy gap. As is sketched in Fig.~\ref{fig:spectrum2dedge}, the size of the band gaps is typically of the order of the hopping amplitude, and therefore it is larger than the bare loss rate $\gamma$, which fixes the characteristic time of the single-resonator dynamics, but still much smaller than the modulation frequency $\Omega_{\rm mod}$. As the same edge state can be probed from different initial sites along the synthetic dimension, the effective operation region consists of a sequence of windows separated by the modulation frequency $\Omega_{\rm mod}$.

\section{4D Quantum Hall effect in a 3D Resonator Lattice}
\label{sec:4D}

To further illustrate the potential of photonic lattices with synthetic dimensions, we now consider a 3D resonator lattice with external modulation, as shown in Fig.~\ref{fig:overview}(c) and described by the tight-binding Hamiltonian $H_J+\sum_\mathbf{r} H_\Omega$ with $j \in {x, y, z}$ [c.f. Eqs.~(\ref{H1site}) and (\ref{eq:hj})]. Including the synthetic dimension, this corresponds to an effective 4D lattice model. As compared to previous works dealing with coupled cavities with higher connectivity~\cite{Jukic:2009,Tsomokos:2010}, our proposal allows for a straightforward inclusion of synthetic gauge fields and is naturally amenable to integration into photonic circuits.

In 4D, the synthetic magnetic field strength can be written as a $4\times4$ antisymmetric tensor with components $B_{k l} \equiv \partial_k A_l - \partial_l A_k$ for each 2D plane given by $k, l \in \{x, y, z, w\}$, where $\mathbf{A}$ is the four-dimensional magnetic vector potential. As discussed in Sec.~\ref{sec:2DA}, the synthetic gauge field is set by the complex hopping phases, where the phases are now indexed by all four lattice coordinates. Just as a synthetic magnetic field in a 2D lattice leads to a 2D quantum Hall effect as shown in Sec.~\ref{sec:2D}, so too can synthetic magnetic fields in a 4D lattice lead to a 4D quantum Hall effect. This effect appears under the application to a filled band of appropriate additional perturbative electric and magnetic fields, denoted here by $E_\nu$ and $\delta B_{\rho \sigma}$ respectively~\cite{IQHE4D, QiZhang, Yang78,Li:2013,Kraus:2013,Price:2015}. 

As shown in Ref.~\cite{Price:2015}, the current in a conservative system in response to these perturbative external fields for a filled lowest (non-degenerate) band has the form:
\begin{align}
	j^\mu
	=
	-e E_\nu \int_{\mathrm{BZ}}\frac{d^4k}{(2\pi)^4} \mathcal{F}^{\mu \nu} (\mathbf{k})
	+
	e^2 E_\nu \delta B_{\rho \sigma} \frac{\nu_2}{(2\pi)^2}\epsilon^{\mu \nu \rho \sigma}, \label{4dcurrent}
\end{align}
with Einstein summation over indices, where $\epsilon^{\mu \nu \rho \sigma}$ is the 4D Levi-Civita symbol and where the integral is over the 4D Brillouin zone (BZ). The Berry curvature $\mathcal{F}^{\mu \nu}$ is defined in terms of the Bloch states of the lowest band $|u_1 (\mathbf{k})\rangle$ as
\begin{align}
	\mathcal{F}^{\mu \nu}
	\equiv
	i
	\left[
	\left\langle \frac{\partial u_1}{\partial k_\mu} \right| \left. \frac{\partial u_1}{\partial k_\nu} \right\rangle
	-
	\left\langle \frac{\partial u_1}{\partial k_\nu} \right| \left. \frac{\partial u_1}{\partial k_\mu} \right\rangle
	\right] , 
\end{align}
and $\nu_2$ is the second Chern number of the first band, defined as
\begin{align}
	\nu_2
	\equiv
	\frac{1}{(2\pi)^2}
	\int_\mathrm{BZ}
	\left( \mathcal{F}^{xy}\mathcal{F}^{zw} + \mathcal{F}^{wx}\mathcal{F}^{zy} + \mathcal{F}^{zx}\mathcal{F}^{yw} \right) d^4 k . 
	\label{secondchern}
\end{align}
The first term in (\ref{4dcurrent}) is a contribution analogous to the quantum Hall effect in 2D systems. This can be seen by noting that this contribution to the current is linear in the applied force $E_\nu$ and that the integral of the Berry curvature over a 2D BZ is related to the first Chern number $\nu_1$. Both of these features are key characteristics of the 2D quantum Hall effect as introduced in Sec.~\ref{sec:2D}. The second term in (\ref{4dcurrent}) is a genuine 4D quantum Hall effect, as can be seen by noting that (i) it is nonlinear in the external fields, (ii) it depends on the second Chern number, and (iii) it vanishes in less than four dimensions due to the Levi-Civita symbol. 

As is evident from the definition of the second Chern number (\ref{secondchern}), a minimal lattice model for the 4D quantum Hall effect requires synthetic magnetic fluxes at least through two ``disconnected" planes.
Along the lines of Refs.~\cite{Kraus:2013,Price:2015}, we consider such a four-dimensional lattice model where the synthetic magnetic field exists in the $xw$ and $yz$ planes. The synthetic gauge field in the $xw$ plane is created in the same way as in Sec.~\ref{sec:2DA}. 

For the flux through the $yz$ plane, we propose to insert non-resonant link resonators in each $xy$ layer with a $z$-dependent lateral displacement [see Fig.~\ref{fig:overview}(c)]. As the resulting hopping phase has a negligible dependence on $w$, this generates the required synthetic gauge field. This approach is inspired by the experiment of Ref.~\cite{Hafezi}, where link resonators were used to engineer a gauge field for photons in a single $xy$ layer, by making the link resonators along $x$ spatially depend on their position along $y$. 

To investigate 4D quantum Hall physics in our synthetic lattice, we generalize the driven-dissipative photonic analog of the 2D integer quantum Hall effect, that was derived in Ref.~\cite{Ozawa:2014} and presented in Sec.~\ref{sec:2Dbulk}, to four dimensions. Here too, by setting the driving frequency on resonance with a band, we can get the required uniform population of a single isolated bulk band, provided that the loss rate satisfies $\Delta \mathcal{E}_{\rm band}<\gamma<\Delta \mathcal{E}_{\rm gap}$. We then consider the response of the system in the $y$ direction under perturbative fields $E_x$ and $\delta B_{zw}$. The perturbative synthetic electric field $E_x$ can be applied by letting the cavity size or the temperature vary uniformly in space. The perturbative magnetic field $\delta B_{zw}$ can be obtained by letting $\mathcal{J}^o_\rr$ have a slow $z$ dependence $\propto e^{i k_z z}$ in addition to the fast one $\propto e^{i k_x x}$ with $k_z \ll k_x$. For simplicity, we take the same hopping amplitude $J$ in all four directions and a uniform loss rate $\gamma$ for all modes.

\begin{figure*}[htbp]
\begin{center}
\includegraphics[width=17cm]{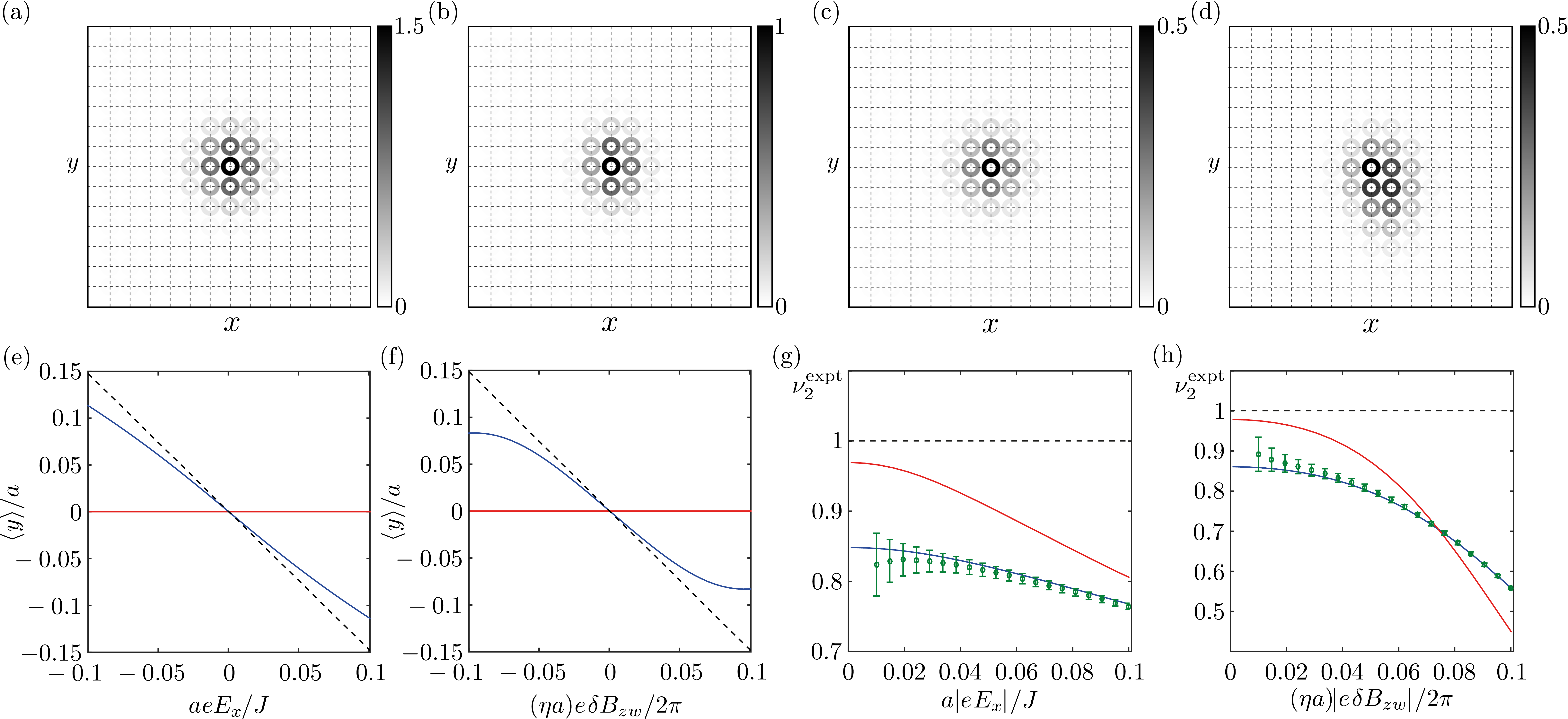}
\caption{(a)-(d) Simulation of photon propagation in the 4D effective lattice sketched in Fig.~\ref{fig:overview}(c), composed of a 3D resonator lattice and one synthetic dimension. The 4D effective lattice has $13^4$ sites, with strong engineered gauge fields in the $xw$ and $yz$ planes of strengths $(\eta a) eB_{xw} = 2\pi/6$ and $a^2 eB_{yz} = 2\pi/7$. The lowest band of this effective 4D topological lattice has a nontrivial second Chern number (\ref{secondchern}), $\nu_2=1$. The central site of the lattice is pumped with frequency $\Omega = -6.3 J$, resonant with the lowest band. We take the uniform loss $\gamma = 0.05J$. The steady-state light intensity distribution is projected onto $xy$ plane for different values of the weak perturbative fields $E_x$ and $\delta B_{zw}$. (a) $a eE_x = 0$ and $(\eta a)e\delta B_{zw} = 0$, (b) $a eE_x = J/10$ and $(\eta a)e\delta B_{zw} = 0$, (c) $a eE_x = 0$ and $(\eta a)e\delta B_{zw} = 2\pi/10$, (d) $a eE_x = J/10$ and $(\eta a)e\delta B_{zw} = 2\pi/10$. The scales show the value of the projected steady-state photon distribution. As can be seen, only in (d) in the presence of both fields is there a clear displacement along $y$: this is the optical analog of the 4D quantum Hall effect. (e) The displacement $\langle y\rangle$ as a function of $E_x$ for a fixed value of $(\eta a) e\delta B_{zw} = 0$ (red horizontal line) and $(\eta a) e\delta B_{zw} = 2\pi/30$ (blue curve). (f) The displacement $\langle y\rangle$ as a function of $\delta B_{zw}$ for a fixed value of $a eE_x = 0$ (red horizontal line) and $a eE_x = J/30$ (blue curve). We take the loss in (e) and (f) to be $\gamma = 0.15J$. The dashed lines in (e) and (f) are the theoretical prediction from Eq. (\ref{4dformula}) without the term $\mathcal{O}(\gamma^0)$ for a band with $\nu_2=1$. (g) and (h) The extracted second Chern number $\nu_2^\mathrm{expt}$ for the configuration of (e) and (f), respectively. The lower (blue) curve at small fields is the estimate using Eq.~(\ref{4dformula}) with one value of loss $\gamma = 0.15J$, ignoring the term $\mathcal{O}(\gamma^0)$. The points with error bars are estimated second Chern numbers for systems with $\pm 10$\% random fluctuations on the loss rate $\gamma_{\mathbf{r},w}$, averaged over 20 different realizations. The upper (red) curve at small fields is the estimate taking into account the term $\mathcal{O}(\gamma^0)$ by using two different values of loss $\gamma = 0.15J$ and $0.16J$ and calculating the difference in the displacement. The true value of the second Chern number $\nu_1 = 1$ is indicated by dashed lines.}
\label{4dsim}
\end{center}
\end{figure*}

In Fig.~\ref{4dsim}, we show the steady-state intensity distribution of photons projected onto the $xy$ plane for a four-dimensional configuration with uniform intrinsic fluxes $(\eta a) eB_{xw} = 2\pi/6$ and $a^2 e B_{yz}= 2\pi/7$. In this case, the lowest band is non-degenerate, extends from $-6.27J$ to $-6.30J$, and is separated from higher-energy bands with a gap equal to $\Delta \mathcal{E}_{\rm gap} = 1.36J$. The driving frequency $\Omega_\mathrm{drive} = -6.30J$ is chosen to be on resonance with the lowest band of second Chern number $\nu_2 = +1$. As expected, only in the presence of both perturbing fields, $E_x$ and $\delta B_{zw}$, is there a clear displacement of the numerically simulated intensity distribution of photons along $y$; this is the hallmark of 4D quantum Hall physics. In addition to this effect, note also the small displacement of the photon intensity distribution also along the electric field direction $x$, which was not predicted by the quantum Hall theory of Eq.~(\ref{4dcurrent}). As for the 2D case, a complete discussion of its origin goes beyond this work and will be given in future work.

We can also quantitatively verify this driven-dissipative analog of 4D quantum Hall physics, by relating the displacement along $y$ to the second Chern number $\nu_2$. For small perturbation fields $E_x$ and $\delta B_{zw}$, it can be shown that (see Appendix~\ref{sec:derivation4d}): 
\begin{eqnarray}
\langle y \rangle = -(\eta a^3)e^2 \frac{q^{yz} q^{xw}}{\gamma} \frac{\nu_2}{(2 \pi)^2} E_x \delta B_{zw} + \mathcal{O}(\gamma^0), \label{4dformula}
\end{eqnarray}
where $q^{yz}$ comes from the flux $e a^2 B_{yz} = 2 \pi (p^{yz}/ q^{yz})$, with $p^{yz}$ and $q^{yz}$ being coprime integers, and similarly for $q^{xw}$. The factor of $(\eta a^3)$ is the volume of a unit cell in the effective four-dimensional lattice. Similarly to Eq.~(\ref{2dphotonformula}), the last term $\mathcal{O}(\gamma^0)$ is a term independent of $\gamma$. The accuracy of this formula is assessed in Figs.~\ref{4dsim}(e) and~\ref{4dsim}(f), where we compare the numerical and analytical predictions as a function of $E_x$ and $\delta B_{zw}$: the analytical prediction is from the formula (\ref{4dformula}) without the term $\mathcal{O}(\gamma^0)$. We have chosen $\gamma = 0.15J$, which covers the lowest band but is much smaller than the band gap. The quite good agreement between the numerical calculation and analytical prediction improves as both perturbative fields, $E_x$ and $\delta B_{zw}$, get weaker.

Reversing the formula (\ref{4dformula}) to estimate $\nu_2$ from the numerical data, one obtains that, ignoring the contribution from the $\mathcal{O}(\gamma^0)$ term, already for significant perturbative fields $a e E_x = J/30$ and $(\eta a) e \delta B_{zw} = 2\pi/30$, the prediction for $\nu_2$ is $0.84$ compared to a true value of $1$. If one eliminates the correction from the $\mathcal{O}(\gamma^0)$ term, the estimate of the second Chern number improves even further. As the term $\mathcal{O}(\gamma^0)$ does not depend on $\gamma$, one can eliminate the contribution from this term by performing simulations (or experiments) with two different values of $\gamma$ and subtracting one from the other (see Sec.~\ref{sec:2Dbulk}). The systematic estimates of the second Chern number as one changes the perturbative fields are plotted in Figs.~\ref{4dsim}(g) and~\ref{4dsim}(h) corresponding to the configurations of Figs.~\ref{4dsim}(e) and~\ref{4dsim}(f), respectively. We plot two curves for Figs.~\ref{4dsim}(g) and~\ref{4dsim}(h), one corresponding to an estimation using one value of loss $\gamma = 0.15J$ without taking into account the term $\mathcal{O}(\gamma^0)$ in the formula (\ref{4dformula}), and the other corresponding to the case of using two values of loss $\gamma = 0.15J$ and $0.16J$ to eliminate the term $\mathcal{O}(\gamma^0)$. We see that at small perturbative fields, using two values of $\gamma$ improves the estimate. We observe in Fig.~\ref{4dsim} (h) that the estimate of the second Chern number is especially sensitive to the increase of $\delta B_{zw}$; a significant deviation appears as soon as the perturbing field $\delta B_{zw}$ becomes comparable to the intrinsic fields $B_{yz}$ and $B_{xw}$ and can no longer be considered as a small perturbation. On the other hand, as shown in Fig.~\ref{4dsim}(g) and~\ref{4dsim}(h), the estimation is only moderately affected by the inclusion of random on-site disorders on the cavity loss rate, $\gamma_{\mathbf r}$.

We have confirmed that the estimate is valid also for other strengths of $B_{yz}$ and $B_{xw}$, as long as the band gap is much larger compared to the bandwidth and the perturbative fields are small enough. These results highlight that an experiment carried out along the lines of our numerical simulation could provide a first experimental measurement of a four-dimensional topological invariant: the emblematic second Chern number.

\section{Discussion}
\label{sec:discussion}

In this section, we now place our proposal in the context of other photonics advances, discussing briefly in turn previous works on synthetic dimensions, on engineered gauge fields and topological edge states in ring-resonator arrays, and on optical isolation in integrated photonics. 

First, there have been two previous proposals that describe how a synthetic dimension may be introduced into a photonics system~\cite{Schmidt:2015, Luo:2015}. In the former~\cite{Schmidt:2015}, the photon and phonon degrees of freedom of optomechanical crystals were utilized giving rise to only two sites in the synthetic dimension, while in the latter~\cite{Luo:2015}, the orbital angular momentum degrees of freedom of optical cavities was exploited as the synthetic dimension with the hopping being provided by spatial light modulators. Compared to these works, our scheme is designed for integrated photonics, and is well suited to applications on chip. We note that while Ref.~\cite{Luo:2015} also studied topological edge states in a 1D array of cavities, the lack of a hard edge in the synthetic dimension meant that light only propagated along the frequency direction. We have instead shown that losses can be used to engineer and control topological edge states along the spatial direction, so that this device could be used as a topological optical isolator. We also emphasise that our proposal allows us to build topological lattice models in dimensions higher than 3D, which may have dramatic applications in integrated photonic circuits with higher connectivity.

Second, our scheme for studying the 2D Harper-Hofstadter model in a 1D chain of multimode resonators can be compared to previous works that implement the model in 2D silicon ring-resonator arrays. In a 2D resonator array, the complex Peierls phase factor is engineered into the spatial coupling between different resonators, for example, using non-resonant link resonators~\cite{Hafezi} (as described also for our 3D resonator lattice in Sec.~\ref{sec:4D}). The use of a synthetic dimension has two key advantages over this experiment; first, the gauge field can be controlled externally through the phase of the dielectric modulation that induces the inter-mode coupling, whereas with link resonators, the gauge field is almost fixed at the time of fabrication with a limited tuning ability offered by a selective local heating of the link resonators~\cite{HafeziCS}. Second, the modulation coupling different modes explicitly breaks time-reversal symmetry. Using link resonators, conversely, this symmetry is not broken as modes with opposite circulation can be viewed as two pseudo-spin experiencing gauge fields of opposite sign. Indeed, in the experiment~\cite{Hafezi}, this meant that the one-way propagation of light was not topologically protected, but could be scattered by pseudo-spin-flipping impurities, such as surface roughness. 

In a similar direction, there has also been a theoretical proposal for engineering the 2D Harper-Hofstadter model in a spatially 2D photonic lattice by modulating the coupling constants between resonators harmonically in time to induce the appropriate Peierls phase factors~\cite{Fan:2012}. While this system shares the advantage of tunability and robustness with our approach, it requires a large real 2D lattice to achieve efficient optical isolation. In our proposal, the left-moving and right-moving states are separated in frequency, and so optical isolation could be realized in a much smaller 1D photonic device. 

Finally, it is instructive to compare our proposal for an optical isolator in Sec.~\ref{sec:2D} to previous works on optical isolation in integrated photonics. Recently, there have been many advances in this field using, for example, magneto-optical elements~\cite{Bi:2011, Tien:2011}, non-linear materials~\cite{Wang:2012, Chang:2014, Peng:2014} or spatiotemporal modulation~\cite{Yu:2009, Lira:2012}. Of these, our set-up is most closely related to those with spatiotemporal modulation, which are also perhaps the most feasible schemes for applications as magneto-optical elements can be difficult to integrate on-chip~\cite{Bi:2011}, while non-linear isolators can be limited, for instance, by dynamic reciprocity~\cite{Shi:2015}. In this approach, spatiotemporal modulations are used to induce indirect interband photonic transitions, for example, in a waveguide on a silicon chip~\cite{Yu:2009,Lira:2012}. By efficiently coupling two photonic modes moving in one direction, but not coupling the modes moving in the opposite direction, such devices break reciprocity and can be used for optical isolation. In a recent experiment, such a modulation was also applied with different phases at two spatially separated points along a waveguide~\cite{Tzuang:2014}; this is conceptually connected to a single plaquette of our synthetic lattice in an engineered magnetic field. By coupling many modes and many ring resonators, our proposal extends such schemes to also exploit the topological robustness of edge states and allows for unidirectional propagation along arbitrary directions in the higher-dimensional space spanned by the physical and synthetic dimensions. This is of utmost interest if one wishes to combine spatial propagation and isolation with a controlled frequency conversion between different channels.

\section{Conclusion}
\label{sec:conclu}

In this paper we have shown how the many-mode nature of ring resonators can be exploited as a synthetic dimension to realize topological photonic lattices on an integrated photonics platform compatible with state-of-the-art silicon photonics technology. We have illustrated our proposal with two important examples, demonstrating its utility and power. We first showed that the 2D quantum Hall effect and its associated topological edge states can be realized using a 1D chain of multimode resonators, which shows a potential to be used as an integrated optical isolator. Second, we demonstrated how the signatures of the 4D quantum Hall effect could be studied in a 3D lattice of resonators.

In addition to being a powerful workhorse for studies of fundamental topological effects in high dimensions, our proposal opens the way towards the realization of photonic circuits with richer connectivities. The synthetic dimension is naturally associated with the different frequency channels of an optical signal, and the topological protection produced by the synthetic gauge field can be exploited to obtain robust unidirectional propagation along specific desired directions in the higher-dimensional space spanned by the physical and synthetic dimensions.

A particularly interesting direction of future research is to explore edge states in four dimensions. The bulk-boundary correspondence tells us that topologically nontrivial bands imply the appearance of topological edge states. The edge of a four-dimensional system should be three-dimensional, with properties that are little known but may lead to interesting applications in higher-dimensional photonic circuits.

In the longer run, one may anticipate that inclusion of strong nonlinearities in such cavity arrays could lead to strongly correlated quantum states of photons that generalize fractional quantum Hall states~\cite{Umucalilar:2012,Umucalilar:2013,Kapit:2014} to high dimensions. In addition to the experimental difficulty of realizing an efficient photon blockade~\cite{carusotto_review} in a multi-resonator and multi-frequency integrated photonics device, this ambitious program raises the further theoretical challenge of finding strategies to obtain fractional quantum Hall states when interactions along the synthetic dimension are non-local and long-ranged, as recently discussed in the cold-atom context in~\cite{Barbarino:2015,Zoller:2015}.

{\it Note added.} Recently, we became aware of a related work by Schleier-Smith on synthetic dimensions in photonics~\cite{Schleier}. Additionally, a paper appeared when a similar concept of synthetic dimensions was proposed in a photonics context focusing on the $1+1$-dimensional case~\cite{Fan}.

\section*{Acknowledgements}

We are grateful to R. Fazio, G. C. La Rocca, M. Ghulinyan, and the members of L. Pavesi's nanolab at University of Trento for continuous exchanges. T.O., H.M.P., and I.C. are supported by the ERC through the QGBE grant, by the EU-FET Proactive grant AQuS, Project No. 640800, and by the Autonomous Province of Trento, partially through the project ``On silicon chip quantum optics for quantum computing and secure communications'' (``SiQuro''). H.M.P was also supported by the EC through the H2020 Marie Sklodowska-Curie Action, Individual Fellowship Grant No. 656093 ``SynOptic". N.G. is financed by the FRS-FNRS Belgium and by the BSPO under PAI Project No. P7/18 DYGEST.  O.Z. acknowledges the Swiss National Foundation for financial support. 

\appendix

\section{Mechanisms for hopping in the synthetic dimension}
\label{sec:implementation}

Having presented relevant examples of topological physics that may be studied with a synthetic dimension in ring resonator arrays, in this appendix we now propose a realistic experimental implementation for how the coupling (\ref{eq:Hmod}) between different optical modes can be achieved. The idea is to externally generate a suitable time-dependent modulation $\delta\chi_{ij}$ of the dielectric tensor of the cavity material so as to coherently convert light from one mode to an adjacent mode. For simplicity, we restrict ourselves here to spatially uniform $\delta\chi_{ij}$ across each resonator. More complex configurations with spatially non-uniform and possibly moving $\delta\chi_{ij}(\mathbf{r},t)$ on each resonator may facilitate an actual implementation of our idea; investigations along these lines will be the subject of future work.

Following quantum nonlinear optics textbooks~\cite{Landau:Book,Butcher:Book,Walls:Book,Drummond:Book}, we denote the electric field profile of the $w$ mode with $\boldsymbol{\mathfrak{E}}^w(\mathbf{r})$ and write the cavity electric field as
\begin{align}
\mathbf{E}(\mathbf{r},t)=\sum_w \boldsymbol{\mathfrak{E}}^{w}(\mathbf{r}) \hat{a}_w + [\boldsymbol{\mathfrak{E}}^{w}(\mathbf{r})]^* \hat{a}^\dagger_w,
\end{align}
with the usual normalization
\begin{align}
	\int d\mathbf{r} \frac{\epsilon (\mathbf{r})}{2\pi}
	[\boldsymbol{\mathfrak{E}}^{w}(\mathbf{r})]^* \cdot \boldsymbol{\mathfrak{E}}^{w^\prime}(\mathbf{r})
	=
	\Omega_w \delta_{w,w^\prime}, \label{normalization}
\end{align}
and $\epsilon (\mathbf{r})$ being the dielectric profile of the cavity.
Only in this section, we take $\mathbf{r}$ to be a continuous variable in space, rather than describing the discrete positions of the resonators as in previous sections.
For modes polarized along the resonator plane (as in the experiment~\cite{Hafezi}), we can approximate the electric index profile in cylindrical $(r,\varphi,z)$ coordinates as 
\begin{align}
 \boldsymbol{\mathfrak{E}}^w(r,\varphi,z)=R_w(r)\,Z_w(z)\,e^{iw\varphi}\,\hat{e}_r, \label{eindex}
\end{align}
with $\hat{e}_r=\cos \varphi\, \hat{e}_x + \sin \varphi\, \hat{e}_y$ defined as the unit vector in the radial direction.

Most generally, the Hamiltonian of the cavity takes the following form under the nonlinearity of the medium 
\begin{align}
	H_\mathrm{mod}
	&=
	-\sum_{i,j}\int d\mathbf{r} \, \delta \chi_{ij}(t)\, E_i (\mathbf{r})\, E_j (\mathbf{r})
	\notag \\
	&\approx
	\sum_{w,w^\prime} H_{w,w^\prime} \hat{a}_w^\dagger \hat{a}_{w^\prime},
	\label{hmod1}
\end{align}
where we have omitted writing out terms, which would become negligible after taking the rotating-wave approximation. The temporally modulated susceptibility tensor $\delta \chi_{ij}(t)$ depends on the pump field used to generate the temporal modulation and the microscopic details of the material we use.
The matrix element of the coupling between a pair of modes $w$ and $w^\prime$ is then
\begin{align}
	H_{w,w'}(t)= -\sum_{i,j}\int d\mathbf{r} \left\{ \delta \chi_{ij}(t) + \delta \chi_{ji}(t)\right\}
	[\mathfrak{E}_i^{w}(\mathbf{r})]^* \mathfrak{E}_j^{w'}(\mathbf{r}). \label{eqm}
\end{align}
As an illustration, we first consider the case where the only non-vanishing entry of $\delta\chi_{ij}$ is a monochromatically oscillating $\delta\chi_{yy}(t)=\delta\bar{\chi}\,\cos(\Omega_{\rm mod} t-\theta)$ with a real $\delta\bar{\chi}$. Restricting to terms that conserve energy in the inter-mode transition, the integral in (\ref{eqm}) then leads for $w>0$ modes to 
\begin{align}
	H_{\mathrm{mod}}
	=\frac{\pi \Omega_w}{2\epsilon} \delta\bar{\chi}  e^{-i \Omega_{\rm mod} t} e^{i \theta}  \sum_{w>0} \hat{a}_{w+2}^\dagger \hat{a}_w  +\textrm{H.c.}, \label{hmodgeneral}
\end{align}
where $\epsilon$ is the dielectric constant.
This expression has the desired form \eq{Hmod} with the identification $\mathcal{J}_{\mathbf{r}}^0 = (\pi \Omega_w / 2\epsilon) \delta \bar{\chi}\,e^{i\theta}$ and a hopping unit $\eta = 2$. In a many-resonator system, a spatially dependent hopping phase and the resulting synthetic gauge field are straightforwardly obtained by letting the modulation phase $\theta$ vary between the different lattice sites.

A straightforward extension of (\ref{hmodgeneral}) to $w<0$ modes shows that modes with the same value of $|w|$ act as two components of a spinor with pseudo-spin $1/2$ and experience the same synthetic gauge field through the Peierls phase-factor $\exp (i \theta)$. Furthermore, as hopping along the $w$ direction is of length $2$, another spin-like degree of freedom is offered by the even or odd parity of the modes. These two families of modes are uncoupled, as a modulation with a different symmetry would be required to couple them, and these again feel the same synthetic gauge field. This crucial difference with respect to the experiment~\cite{Hafezi} is extremely beneficial for applications as it guarantees an improved robustness of the unidirectional edge states.

We now present three specific experimental configurations that can lead to the desired modulation of the dielectric tensor. In all these cases, the modulation is introduced through an externally imposed, spatially varying electromagnetic field, and the efficiency of the resulting hopping is quantified by the ratio of the dielectric index shift $\delta\bar{\chi}$ over the resonator quality factor $\Omega_w / \gamma$.

\subsection{Quasi-static $\chi^{(3)}$ nonlinearity}

We start by considering the case where the underlying medium has a $\chi^{(3)}$ nonlinearity and the modulation frequency $\Omega_{\rm mod}$ lies well below the optical gap of the material, so that we can approximate the nonlinear response as quasi-static. Assuming that the material has cubic symmetry, the time-dependent modulation of the dielectric constant takes the following form under the influence of the pump beam $\mathbf{E}^P (t)$:
\begin{equation}
 \delta \chi_{ij}(t)=\bar{\chi}^{(3)}\,\left( \alpha E^{P}_i(t)\,E^{P}_j(t) + \beta |\mathbf{E}^P(t)|^2\,\delta_{ij}\right), \label{eq:ins}
\end{equation}
where the coefficients $\alpha$ and $\beta$ are determined from the microscopic details of the material.
(For materials with different symmetries, similar discussions follow {\it mutatis mutandis}.)
Using the ansatz (\ref{eindex}) and the normalization (\ref{normalization}), the integral in (\ref{eqm}) can be performed to find the Hamiltonian
\begin{align}
	&H_\text{mod}
	=
	-\frac{2\pi \Omega_{w_0} \bar{\chi}^{(3)}}{\epsilon}
	\sum_w
	\left(
	(\alpha + 2\beta)\left| \mathbf{E}^P\right|^2 \hat{a}_w^\dagger \hat{a}_w
	\right.
	\notag \\
	&
	\left.
	+\alpha \left( \left(E_x^{P}\right)^2 - \left( E_y^{P}\right)^2 - 2iE_x^P E_y^P \right)
	\hat{a}^\dagger_{w+2} \hat{a}_w
	\right)
	+\text{H.c.}, \label{hammod}
\end{align}
where we have assumed that the cavity electric field is well confined in the medium so that the dielectric constant $\epsilon$ does not spatially vary, and also that $w_0 \gg 0$ so that $| \mathbf{\mathcal{E}}^w | \approx | \mathbf{\mathcal{E}}^{w_0} |$.
For a $y$-polarized monochromatic pump beam with frequency $\Omega_\mathrm{mod}/2$ and (complex) amplitude $\bar{E}^P$
\begin{align}
	\mathbf{E}^P (t)
	=
	\mathrm{Re}\left[ \bar{E}^P e^{-i\Omega_\mathrm{mod} t/2} \hat{e}_y \right],
\end{align}
one has $E_x^P = 0$ and
\begin{multline}
	\left( E_y^P(t) \right)^2
	=
	\frac{1}{4}\left( \left( \bar{E}^{P} \right)^2  e^{-i\Omega_\text{mod}t} +\right. \\ \left. + \left( \bar{E}^{P *}\right)^2 e^{i\Omega_\text{mod} t} + 2 \left| \bar{E}^P \right|^2 \right).
\end{multline}
Inserting this expression in the tight-binding Hamiltonian (\ref{hammod}), one obtains
\begin{align}
	H_\text{mod}
	\approx
	&\frac{\pi \Omega_{w_0} \bar{\chi}^{(3)}}{\epsilon}\alpha \left( \bar{E}^{P} \right)^2  e^{-i\Omega_\text{mod}t}
	\notag \\
	&\sum_w
	\left(
	\hat{a}^\dagger_{w+2} \hat{a}_w + \hat{a}^\dagger_{w-2} \hat{a}_w
	\right)
	+\text{H.c.}, \label{chi3ham}
\end{align}
where we have omitted static terms proportional to $\hat{a}_w^\dagger \hat{a}_w$, which just shift the on-site energy uniformly, as well as terms that can be neglected after the rotating-wave-approximation.
As we can see, the oscillating terms proportional to $\hat{a}_{w+2}^\dagger \hat{a}_w$ indeed provide the desired hopping for $w > 0$ with $\eta = +2$ in the synthetic dimension as posed in~\eq{Hmod}. The terms proportional to $\hat{a}_{w-2}^\dagger \hat{a}_w$ provide hopping for $w < 0$ with $\eta = -2$.

Note that the amplitude of these hopping terms is proportional to the square of the pump field amplitude $\left( \bar{E}^P \right)^2$, which then controls the hopping phase: the underlying physical process is in fact the conversion of a resonator photon in the mode at $\Omega_w$ into a resonator photon in the mode at $\Omega_{w+2}$ by mixing with two photons of the modulation field at $\Omega_\textrm{mod}/2$. As a result, if one has a chain of resonators along the $x$ direction, one can spatially vary the complex hopping phase by applying a plane-wave beam along the $x$ direction so that $\bar{E}^P \propto e^{ikx}$. We also note that the amplitude of the hopping is $\sim \Omega_{w_0} \bar{\chi}^{(3)}$ times the intensity of the applied pump laser. 
As discussed in the end of Sec.~\ref{sec:2DA}, for the practical purpose one needs to have $|\mathcal{J}_\mathbf{r}^0|/\Omega_{w_0} > 1/Q$. Using (\ref{chi3ham}), this condition translates to $\bar{\chi}^{(3)}\alpha \left( \bar{E}^{P} \right)^2 \gtrsim 1/Q$, where the left-hand side is the dielectric modulation applied to the system. If we naively take values for silicon, $\chi^{(3)} \sim 10^{-19} (\text{m/V})^2$, with a $Q$ value of $\sim 10^5$, one needs an electric field of $\bar{E}^{P} \sim 10^7 \text{V/m}$, which is not practical. One can, however, use more sophisticated configurations to obtain larger modulations. For example, in Ref.~\cite{Lira:2012}, it is argued that using $pn$ junctions as the nonlinear element one can achieve the dielectric modulation as high as $\sim 0.23$ in the THz range for realistic electric field intensities.

\subsection{$\chi^{(3)}$ nonlinearity in a Raman-like scheme}

In order to take advantage of the typically stronger value of the $\chi^{(3)}$ nonlinear susceptibility at optical frequencies, we can adopt a different scheme using a two-frequency pump with components at $\Omega_P + \Omega_\mathrm{mod}$ and $\Omega_P$, whose frequency difference is tuned to the desired modulation frequency according to a Raman-like scheme. The total pump field, polarized along $y$ as before, is then
\begin{align}
	\mathbf{E}^P (t)
	=
	\mathrm{Re}\left[ \left( \bar{E}^P_1 e^{-i\Omega_\mathrm{mod} t} + \bar{E}^P_2 \right) e^{-i\Omega_P t} \hat{e}_y \right].
\end{align}
Assuming that the carrier $\Omega_P \gg \Omega_\mathrm{mod}$, we can perform a rotating-wave approximation neglecting polarization terms at frequencies of the order of $\pm 2\Omega_P$.
The effective dielectric modulation seen by a low-frequency beam is
\begin{equation}
 \delta \chi_{yy}(t)=\alpha\, \bar{\chi}^{(3)}\,\left| \bar{E}^P_1 e^{-i\Omega_\mathrm{mod} t} + \bar{E}^P_2 \right|^2 , \label{eq:ins2}
\end{equation}
where the value of $\bar{\chi}^{(3)}$ is to be evaluated for the Raman process under investigation. Here, we have neglected the rotationally symmetric component of the dielectric modulation $\delta\chi_{ij}$ proportional to $\beta$, which in the end just shifts the overall frequency of each cavity but does not affect the physics. The intensity 
\begin{align}
&\left| \bar{E}^P_1 e^{-i\Omega_\mathrm{mod} t} + \bar{E}^P_2 \right|^2 \label{Ep2} \\
&=|\bar{E}^P_1|^2 + |\bar{E}^P_2|^2 +\bar{E}^{P*}_2 E^P_1 \,e^{-i\Omega_\mathrm{mod} t} + \bar{E}^{P}_2 \bar{E}^{P*}_1 \,e^{i\Omega_\mathrm{mod} t}, \notag
\end{align}
is modulated at $\Omega_\textrm{mod}$ as a result of the interference of the two frequency components.
Including this expression into the dielectric modulation (\ref{eq:ins2}) and then into the tight-binding Hamiltonian (\ref{eqm}) and (\ref{hmodgeneral}), one obtains the following Hamiltonian
\begin{align}
	H_\mathrm{mod}
	=&
	\frac{\pi \alpha \bar{\chi}^{(3)} \Omega_w}{\epsilon}
	\bar{E}^{P*}_2 \bar{E}^P_1 \,e^{-i\Omega_\mathrm{mod} t}
	\notag \\
	&\sum_w
	\left(
	\hat{a}_{w+2}^\dagger \hat{a}_w + \hat{a}_{w-2}^\dagger \hat{a}_w
	\right)
	+\text{H.c.},
\end{align}
where, as in the previous subsection, we only kept terms that are relevant after the rotating-wave approximation.

Physically, one can understand the process as follows: the pump beam excites the material proportionally to its instantaneous intensity $\left| \bar{E}^P_1 e^{-i\Omega_\mathrm{mod} t} + \bar{E}^P_2 \right|^2$ and the modulation of the dielectric constant instantaneously follows the excitation. Alternatively, one can think of a pump photon of the $\Omega_P+\Omega_\textrm{mod}$ component and a resonator photon in the mode at $\Omega_w$ being converted into an extra pump photon at the other frequency $\Omega_P$ plus a resonator photon in the mode at $\Omega_{w+2}$.

The phase of the hopping in the synthetic dimension is now controlled by the relative phase of the two frequency components, that is, by the argument of $\bar{E}_1^P \bar{E}_2^{P *}$.
A main advantage of this scheme is that the $\chi^{(3)}$ nonlinear optical susceptibility involved in the process is evaluated for a carrier frequency $\Omega_P$ in the optical domain. As $\chi^{(3)}$ in this domain can be made orders of magnitude larger by approaching and possibly entering the band gap of the material~\cite{Butcher:Book}, the required power in the pump beams is much less demanding. A potentially useful further reinforcement of $\chi^{(3)}$ can be induced by tuning the modulation on resonance with an optical phonon mode of the material (in the 15~THz range in Si)~\cite{Petek:2012}.

\subsection{Quasi-static $\chi^{(2)}$ nonlinearity}

As a final example, it is also possible to use a $\chi^{(2)}$ Pockels electro-optic modulation of the dielectric tensor, although this requires the use of a non-centro-symmetric cavity material, which is not the case in standard silicon devices. To achieve this, one can include other materials in the device, or, alternatively, recent works have suggested that a suitable mechanical strain might induce a strong $\chi^{(2)}$ even in pure silicon~\cite{Pavesi:2012}. For the purposes of this illustrative discussion, we assume the crystal structure has a trigonal symmetry described by a point group $3m$, such as LiNbO$_3$. Assuming that both the cavity electric field and the pump field do not have any $z$ component, the nonlinear susceptibility tensor $\chi^{(2)}_{\mu \alpha \beta}$ then has only one independent component:
\begin{align}
	\bar{\chi}^{(2)}
	\equiv
	\chi^{(2)}_{xxy}
	=
	\chi^{(2)}_{xyx}
	=
	\chi^{(2)}_{yxx}
	=
	-\chi^{(2)}_{yyy}.
\end{align}
This implies that, when the cavity is irradiated with a pump field $\mathbf{E}^P(t)$, the quasi-static modulation of the dielectric constant via the Pockels effect has the form
\begin{align}
	\delta \chi_{xx}
	&=
	-\delta \chi_{yy}
	=
	\bar{\chi}^{(2)}E^P_y,
	\notag \\
	\delta \chi_{xy}
	&=
	\delta \chi_{yx}
	=
	\bar{\chi}^{(2)} E^P_x.
\end{align}
The integral in (\ref{eqm}) then yields the Hamiltonian
\begin{align}
	H_{\mathrm{mod}}
	=
	-\frac{2\pi\Omega_{w_0} \bar{\chi}^{(2)}}{\epsilon}\sum_w
	(E_y^P - iE_x^P)\hat{a}_{w+2}^\dagger \hat{a}_w + \text{H.c.}. \label{hmodchi2}
\end{align}
As before, we apply a pump field of frequency $\Omega_\mathrm{mod}$ and (complex) amplitude $\bar{E}^P$
\begin{align}
	\mathbf{E}^P(t)
	=
	\mathrm{Re}
	\left[
	\bar{E}^P e^{-i\Omega_{\mathrm{mod}}t}
	\hat{e}_y
	\right], \label{pumpfield}
\end{align}
with a linear polarization along $\hat{e}_{y}$. Inserting this expression into (\ref{hmodchi2}), one obtains
\begin{align}
	&H_{\mathrm{mod}}
	=
	-\frac{\pi\Omega_{w_0}\bar{\chi}^{(2)}}{\epsilon}\bar{E}^P
	\notag \\
	&\times \sum_w	
	\left( \hat{a}_{w+2}^\dagger \hat{a}_w + \hat{a}_{w-2}^\dagger \hat{a}_{w} \right)
	e^{-i\Omega_\mathrm{mod}t}
	+ \text{H.c.}, \label{eq:Hmodf}
\end{align}
which has the desired form with the identification
$\mathcal{J}_{\mathbf{r}}^0 = (\pi \Omega_{w_0}\bar{\chi}^{(2)}/\epsilon) \bar{E}^P$
again with the hopping units $\eta = \pm 2$. 

In this scheme, the hopping phase is determined by the phase of $\bar{E}^P$: the underlying physical process is in fact the conversion of a resonator photon in the mode at $\Omega_w$ into a resonator photon in the mode at $\Omega_{w+2}$ by mixing with one photon of the modulation field at $\Omega_\textrm{mod}$. Note that the missing angular momentum is provided by the anisotropy of the considered $\chi^{(2)}$ material.

\section{Analytical model for the spatial displacement of the intensity distribution}
\label{sec:analyticalmodel}

We present a simple analytical model that shows how the maximum of the light intensity injected into a unidirectionally propagating mode is significantly shifted away from the peak of the driving field intensity towards the downstream direction. Although this model differs in a few aspects from the lattice model considered in the main text, it captures some of the general features of unidirectional light propagation.

We consider a continuous-space, one-dimensional field model with a unidirectionally propagating dispersion centered around the frequency $\omega_0$ as
\begin{equation}
 \omega(k)=v_{\rm gr}\,k + \omega_0,
\end{equation}
and a loss rate $\gamma$, described by an equation of motion of the form
\begin{equation}
 i\frac{\partial\psi}{\partial t}= -i v_{\rm gr}\frac{\partial\psi}{\partial x} + (\omega_0 - i\gamma) \psi + F(x,t),
\end{equation}
where the coherent driving field $F(x,t)=F(x)\,e^{-i\Omega_{\rm drive} t}$ is assumed to be monochromatic at frequency $\Omega_{\rm drive}$ with an exponential spatial profile of small size $\sigma$,
\begin{equation}
 F(x)=F_0\,e^{-|x|/\sigma},
 \label{eq:Fx}
\end{equation}
whose spatial Fourier transform is
\begin{equation}
 \tilde{F}(k)=\frac{2F_0}{\sigma}\frac{1}{k^2+\frac{1}{\sigma^2}}.
\end{equation}
This specific shape was chosen for analytical convenience, but we have numerically checked that the same conclusions hold for other localized shapes of the driving field.

The steady state of the field has the form $\psi(x,t)=\psi(x)\,e^{-i\Omega_{\rm drive} t}$, which can be found by solving the time evolution in Fourier space, giving
\begin{equation}
 \psi(k)=-\frac{\tilde{F}(k)}{\omega(k)-\Omega_{\rm drive} + \omega_0 -i\gamma}.
\end{equation}
Thanks to our specific choice for the driving field profile, this result can be easily Fourier-transformed back to the real-space field, getting:
\begin{equation}
 \psi(x)=F_0 \left[A_+(x) \Theta(x) + A_-(x) \Theta(-x)\right],
\end{equation}
with
\begin{align}
 A_-(x)&= \frac{e^{x/\sigma}}{\Omega_{\rm drive}-\omega_0+i\gamma+i\frac{v_{\rm gr}}{\sigma}}, \notag \\
 A_+(x)&= \frac{e^{-x/\sigma}}{\Omega_{\rm drive}-\omega_0+i\gamma-i\frac{v_{\rm gr}}{\sigma}} \\
 &- 
 \frac{2i v_{\rm gr}/\sigma}{(\Omega_{\rm drive}-\omega_0+i\gamma)^2+\frac{v_{\rm gr}^2}{\sigma^2}}
 e^{i(\Omega_{\rm drive} - \omega_0 + i\gamma) x/v_{\rm gr}}, \notag
\end{align}
where $\Theta (x)$ is the Heaviside step function. The second term of $A_+$ describes the radiative wave that is injected by the driving field into the cavity and propagates in the positive $x$ direction with a wave vector $(\Omega_{\rm drive} - \omega_0)/v_{\rm gr}$, which is determined by the resonance condition between the driving field and the cavity mode. The absence of a similar wave in the negative $x$ direction is a signature of the unidirectional nature of the propagation. The spatial decay of this wave occurs on a long length scale proportional to the absorption length $\ell_{\rm abs}=v_{\rm gr}/2\gamma$.
Near the driven region, the field amplitude contains an additional, localized contribution that spatially decays on a much shorter length scale fixed by the size of the spot $\sigma$. The superposition of the different contributions provides the smooth transition connecting the two sides of the driven region.

\begin{figure}[htbp]
 \includegraphics[width=0.95\columnwidth]{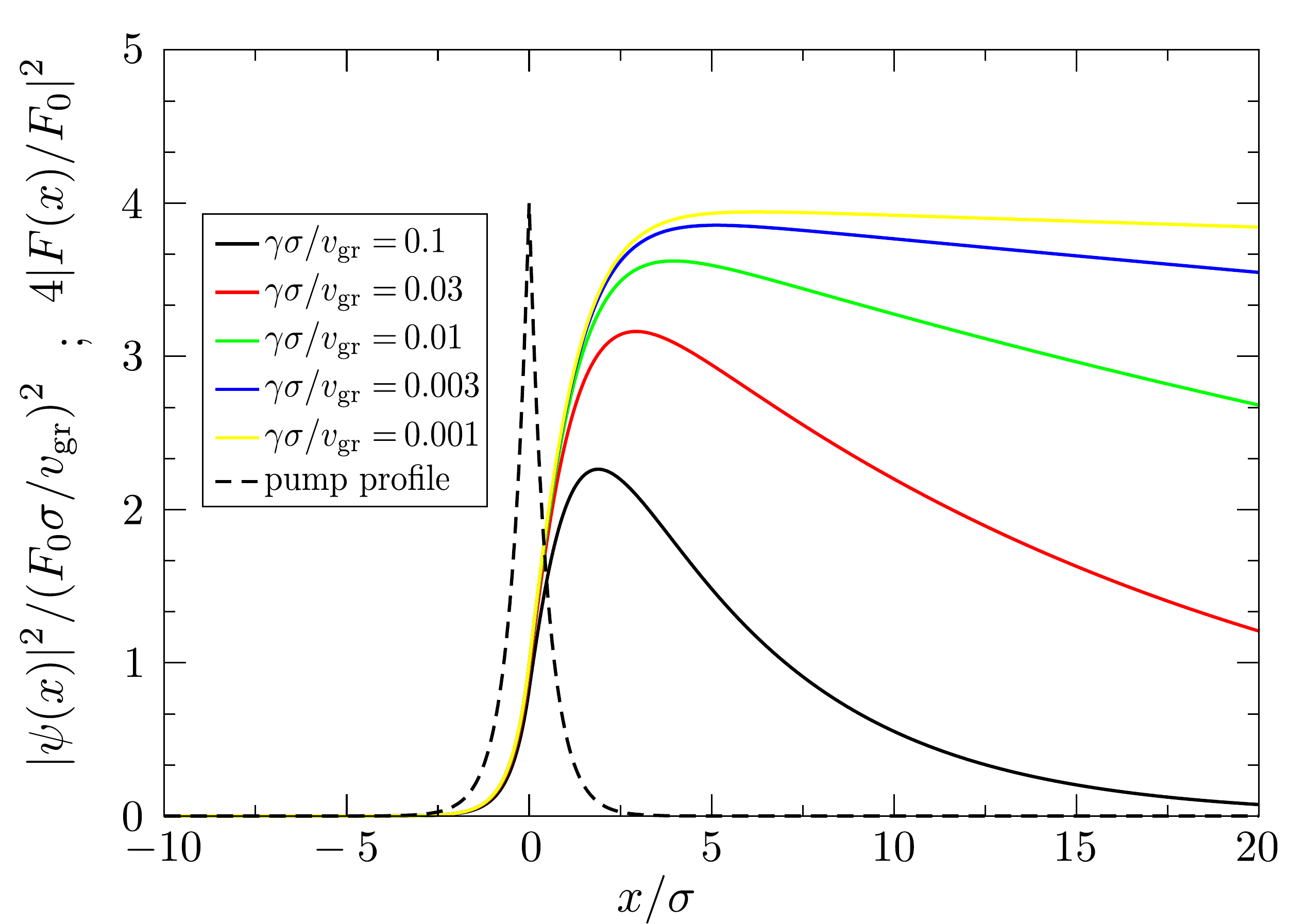}
\caption{The solid lines show the spatial profile of the light intensity $|\psi( x )|^2$, in units of $(F_0 \sigma / v_\mathrm{vg})^2$, in a unidirectionally propagating mode of a cavity. The driving field $F(x)$ is monochromatic, satisfying the resonance condition $\Omega_\mathrm{drive} = \omega_0$ with an exponentially shaped spot. The dashed line plots the intensity of the driving field scaled as $4|F(x)/F_0|^2 $. The different solid curves correspond to different values of the cavity decay as indicated in the legend. }
\label{fig:unidir_peak}
\end{figure}

As one can see in Fig.\ref{fig:unidir_peak}, this superposition determines the position of the field intensity maximum, which turns out to be displaced from the driving center by an amount that decreases for increasing loss rate $\gamma$ (i.e., a decreasing absorption length $\ell_{\rm abs}$ at a given $v_{\rm gr}$). This prediction of the simple analytical model provides a qualitative explanation as to what was observed in the full calculations presented in Figs.~\ref{fig:2dedge} and~\ref{fig:transmission} of the main text. 

As already mentioned, even though we have presented calculations for a very specific choice of the driving field profile for which the analytical calculation is the simplest, we have numerically checked that the same conclusions hold, e.g., for Gaussian-shaped profiles. Even a lattice model of spacing $a$ can be mimicked in our continuum model by restricting the Fourier space to $-\pi/a<k<\pi/a$: also in this case, our conclusions about the spatial displacement remain valid.

\section{Derivation of the driven-dissipative 4D quantum Hall formula}
\label{sec:derivation4d}

We derive the formula Eq.~(\ref{4dformula}). A similar formula was derived in Ref.~\cite{Ozawa:2014} for the 2D quantum Hall effect, and we extend the formula to 4D. In Sec.~IV, we separated magnetic fields into intrinsic field and perturbative field. The strategy we use here is to first treat all the magnetic fields as intrinsic and derive the formula for the lateral displacement: up to this point, the formula will have the same form as 2D. Then, we separate the magnetic field into intrinsic portion and perturbative portion and derive how the response depends on the second Chern number.

We consider a lattice model where bands have nontrivial topology. We drive the system with frequency $\Omega_\mathrm{drive}$ around the center of the lattice so that we can ignore edge effects. Assuming that all the magnetic fields are intrinsic, the only perturbation is the electric field, and in this case, the formula
\begin{align}
	\langle x^\mu\rangle
	=
	-\frac{e E_\nu}{\gamma}
	\frac{\displaystyle \sum_{n \in \mathcal{B}}\int_{\overline{\mathrm{BZ}}} \frac{d^d k}{(2\pi)^d} \bar{\mathcal{F}}_n^{\mu \nu}(\mathbf{k})}{\displaystyle \sum_{n \in \mathcal{B}} \int_{\overline{\mathrm{BZ}}}\frac{d^d k}{(2\pi)^d}},
	\label{included}
\end{align}
derived in Ref.~\cite{Ozawa:2014} holds irrespective of the dimensionality.
Here $\bar{\mathcal{F}}_n^{\mu \nu}(\mathbf{k})$ is the Berry curvature in the $\mu \nu$ plane, when we regard all the magnetic fields in the system as intrinsic, and $\overline{\mathrm{BZ}}$ is the corresponding Brillouin zone in such a case. We assume that the loss $\gamma$ is larger than the bandwidth of a set of closely spaced bands $\mathcal{B}$, but smaller than the rest of the bands.

The numerator of~(\ref{included}) is the anomalous current in the presence of the perturbative synthetic electric field (Eq. (5) of Ref.~\cite{Price:2015} when the perturbative magnetic field is zero).
We now separate the magnetic fields into an intrinsic contribution and a perturbative part. From Ref.~\cite{Price:2015}, we find the following relation for the current:
\begin{align}
	j^\mu
	&=
	-e E_\nu \sum_{n \in \mathcal{B}} \int_{\overline{\mathrm{BZ}}}\frac{d^dk}{(2\pi)^4} \bar{\mathcal{F}}^{\mu \nu}_n (\mathbf{k})
	\notag \\
	&=
	-e E_\nu \int_{\mathrm{MBZ}}\frac{d^dk}{(2\pi)^4} \mathcal{F}^{\mu \nu} (\mathbf{k})
	+
	e^2 E_\nu \delta B_{\rho \sigma} \frac{\nu_2}{(2\pi)^2}\epsilon^{\mu \nu \rho \sigma},
\end{align}
where $\mathcal{F}^{\mu \nu}$ is the Berry curvature of the occupied band and $\mathrm{MBZ}$ is the magnetic Brillouin zone in the absence of the perturbative synthetic electric and magnetic fields.
We have assumed that only one band is occupied in the absence of the perturbative fields. (Extension to a more general situation is trivial.)
This equation allows us to write the numerator of (\ref{included}) in terms of the intrinsic quantities.
On the other hand, the denominator of (\ref{included}) is just the total particle density and thus~\cite{Price:2015}
\begin{align}
	\sum_{n \in \mathcal{B}} \int_{\overline{\mathrm{BZ}}}\frac{d^d k}{(2\pi)^d}
	=
	\int_{\mathrm{MBZ}} d^d k D(\mathbf{k}),
\end{align}
where $D(\mathbf{k})$ is the modified phase-space density of state:
\begin{align}
	D(\mathbf{k})
	=&
	\frac{1}{(2\pi)^4}
	\left(
	1 + \frac{1}{2}\delta B_{\rho \sigma}\mathcal{F}^{\rho \sigma}
	\right.
	\notag \\
	&\left.
	+
	\frac{1}{64}
	\left( \varepsilon^{\mu \nu \rho \sigma} \delta B_{\mu \nu} \delta B_{\rho \sigma} \right)
	\left( \varepsilon_{\mu \nu \rho \sigma} \mathcal{F}^{\mu \nu} \mathcal{F}^{\rho \sigma} \right)
	\right).
\end{align}
Combining everything together in four dimensions, (\ref{included}) is
\begin{align}
	\langle x^\mu\rangle
	=
	\frac{\displaystyle -e E_\nu \int_{\mathrm{MBZ}}\frac{d^4k}{(2\pi)^4} \mathcal{F}^{\mu \nu} (\mathbf{k})
	+
	e^2 E_\nu \delta B_{\rho \sigma} \frac{\nu_2}{(2\pi)^2}\epsilon^{\mu \nu \rho \sigma}}
	{\displaystyle \gamma \int_{\mathrm{MBZ}} d^4 k D(\mathbf{k})}.
\end{align}
In the configuration we consider in Sec.~IV, the first term in the numerator is zero. At the same time, for this configuration the modified phase-space density of state is just $D = 1/(2\pi)^4$, and thus the denominator is $\gamma / ( \eta a^3 q^{yz} q^{xw})$. Combining these facts, we immediately obtain Eq.~(\ref{4dformula}). A complete discussion of the subtle features relating the spatial shift of the center of mass of the intensity distribution to the more usual concept of Hall current can be found in our recent paper~\cite{MagnPert}.

\end{document}